\documentclass[a4paper,11pt]{article}
\pdfoutput=1 
\usepackage{jcappub}
\usepackage{amsmath}  
\usepackage{dsfont} 
\usepackage{comment}
\usepackage[T1]{fontenc} 
\usepackage{enumerate}
\usepackage[color=orange]{todonotes}
\usepackage[normalem]{ulem}

 \usepackage[normalem]{ulem}
 \usepackage{hyperref}
\usepackage{academicons}
\usepackage{xcolor}
\usepackage{orcidlink}
\usepackage{changepage}

\newcommand{\orcid}[1]{\href{https://orcid.org/#1}{\textcolor[HTML]{A6CE39}{\aiOrcid}}}

\newcommand{\CL}{{\tt ${\mathcal C}$osmo${\mathcal L}$attice}~}

\title{\begin{adjustwidth}{0cm}{-3cm} \boldmath \fontsize{19}{20}\selectfont Ephemeral Oscillons in Scalar-Tensor Theories: \\ The Higgs-like case\end{adjustwidth}
}

\author[1]{Matteo Piani,}
\author[2]{Javier Rubio,}
\author[3]{Francisco Torrenti}
\affiliation[1]{Centro de Astrof\'{\i}sica e Gravita\c c\~ao  - CENTRA,
Departamento de F\'{\i}sica, Instituto Superior T\'ecnico - IST,
Universidade de Lisboa - UL, Av. Rovisco Pais 1, 1049-001 Lisboa, Portugal.}
\affiliation[2]{\,\,Departamento de Física Teórica and Instituto de Física de Partículas y del Cosmos (IPARCOS-UCM), Universidad Complutense de Madrid, 28040 Madrid, Spain}
\affiliation[3]{Departament de Física Quàntica i Astrofísıca \& Institut de Ciències del Cosmos (ICCUB),
Universitat de Barcelona, Martí i Franquès 1, 08028 Barcelona, Spain}

\emailAdd{matteo.piani@tecnico.ulisboa.pt}
\emailAdd{javier.rubio@ucm.es}
\emailAdd{f.torrenti@ub.edu}
\abstract{We investigate the post-inflationary evolution of a non-minimally coupled inflaton field in scalar-tensor theories, framed within the flexible framework of Einstein-Cartan gravity. By focusing on a class of simplified Higgs-like scenarios, we simulate the transition from the end of inflation to the formation of oscillons using fully-fledged 3+1 classical lattice simulations. Once oscillons have formed, we extract their profiles and perform 1+1 simulations to evolve their radial equations. Our findings reveal that these oscillons, unlike typical cases in the literature, are relatively short-lived, due to the presence of self interactions at small field values. The radiation produced by this novel type of oscillons can quickly lead to a radiation-dominated Universe, even in the absence of additional fields or interactions. Finally, we leverage our results to derive precise predictions for the inflationary observables and the produced spectrum of gravitational waves associated with oscillon formation. Significantly, our study establishes also an upper bound on the duration of the heating phase in Einstein-Cartan Higgs inflation scenarios.}

\begin{document}
%%%%%%%%%%%%%%%%%%%%%%%%%%%%%%%%%%%%%%%%%%%%%%%%%%%%%%%%%%%%%%%%%%%%%%%%%%%%%%%%
\maketitle
\flushbottom

%%%%%%%%%%%%%%%%%%%%%%%%%%%%%%%%%%%%%%%%%%%%%%%%%%%%%%%%%%%%%%%%%%%%%%%%
\section{Introduction}\label{sec:Intro}
%%%%%%%%%%%%%%%%%%%%%%%%%%%%%%%%%%%%%%%%%%%%%%%%%%%%%%%%%%%%%%%%%%%%%%%%

Precision measurements of the Cosmic Microwave Background (CMB) \cite{Planck:2018jri, BICEP:2021xfz} have firmly established inflation \cite{Guth:1980zm, Linde:1981mu, Mukhanov:1981xt} as the standard paradigm to address the shortcomings of the hot Big Bang model while providing a natural mechanism to generate the primordial density perturbations seeding structure formation.

In its simplest form, inflation is driven by a homogeneous scalar field, the inflaton, which, slowly rolling down a potential, is capable of supporting a sufficiently long period of accelerated expansion. However, the identity of the inflaton field remains elusive, leading to two interrelated challenges in the construction of inflationary models. First, the shape of the inflationary potential must comply with precise observational constraints, such as the mild scale dependence of the power spectrum of primordial density perturbations and the increasingly tightening limits on the tensor-to-scalar ratio, encoding the amplitude of the so far undiscovered primordial gravitational waves background~\cite{BICEP:2021xfz}. Second, inflation must gracefully transition from a quasi-de Sitter expansion epoch to a decelerating \textit{heating} phase, when the Universe becomes radiation dominated and eventually reaches thermal equilibrium \cite{Bassett:2005xm,Allahverdi:2010xz, Amin:2014eta}. Understanding this intermediate stage is crucial for setting the initial conditions for Big Bang Nucleosynthesis (BBN) \cite{Kawasaki:1999na,Kawasaki:2000en,Hannestad:2004px,Hasegawa:2019jsa} and determining the precise number of $e$-folds entering the key inflationary observables \cite{Dai:2014jja,Martin:2014nya,Munoz:2014eqa,Gong:2015qha,Cook:2015vqa}. To further complicate this picture, the so-called \textit{preheating} epoch or initial stage of heating \cite{Traschen:1990sw,Kofman:1994rk,Shtanov:1994ce,Kaiser:1995fb,Khlebnikov:1996mc,Prokopec:1996rr,Khlebnikov:1996wr,Kaiser:1997mp,Kofman:1997yn,Greene:1997fu,Khlebnikov:1996zt,Kaiser:1997hg} is typically highly non-linear and non-perturbative, customarily requiring the use of numerical methods such as classical lattice simulations \cite{Figueroa:2020rrl}.

A fascinating byproduct of inflaton fragmentation during preheating in theories with potentials shallower than quadratic is the formation of localized quasiperiodic scalar field configurations called oscillons \cite{Bogolyubsky:1976yu,Gleiser:1993pt,Copeland:1995fq,Amin:2010dc,Amin:2011hj,Antusch:2015nla,Lozanov:2017hjm,Hasegawa:2017iay,Antusch:2017flz,Sang:2019ndv,Antusch:2019qrr,Ibe:2019lzv,Kou:2019bbc,Sang:2020kpd,Aurrekoetxea:2023jwd,Mahbub:2023faw,vanDissel:2023zva}. These nonlinear structures have drawn significant attention for their potential to dominate the post-inflationary expansion of the Universe, inducing a prolonged early matter-dominated era.  Interestingly enough,  they can also act as significant sources of stochastic gravitational wave backgrounds (SGWB) \cite{Zhou:2013tsa,Antusch:2016con,Antusch:2017vga,Amin:2018xfe,Liu:2018rrt,Lozanov:2019ylm,Lozanov:2022yoy,Piani:2023aof}, beyond the frequency reach of current detectors but within the sensitivity range of proposed ultra-high frequency detectors \cite{Aggarwal:2020olq,Herman:2022fau,Aggarwal:2025noe}.
Although extensively studied in single-field inflationary models —ranging from polynomial potentials \cite{Antusch:2015nla,Antusch:2019qrr,Drees:2025iue} to $\alpha$-attractor frameworks \cite{Mahbub:2023faw,Lozanov:2017hjm,Lozanov:2019ylm} or axion monodromy potentials \cite{Amin:2011hj,Zhou:2013tsa,Lozanov:2019ylm}—, most research on oscillon formation and evolution during preheating has largely neglected the role of inflaton interactions with external fields and gravity, with the few exceptions in the literature revealing intricate patterns of formation, stability, and decay \cite{Antusch:2015ziz,Shafi:2024jig}.

In this work, we explore the formation and decay of oscillons within scalar-tensor theories inspired by the Einstein-Cartan (EC) formulation of gravity. By extending general relativity to incorporate torsion as a dynamical quantity, these theories provide a rather general geometrical framework for describing the interplay between scalar fields and gravity, while encompassing both metric and Palatini formulations as limiting cases. Specifically, we focus on Jordan-frame formulations featuring polynomial operators of dimension $d\leq 4$, with at most two field derivatives and a purely quartic, scale-invariant potential mimicking the behavior of the Standard Model Higgs one at large field values. This set of requirements leads to an effective Einstein-frame inflationary potential characterized by an asymptotic plateau at super-Planckian field values and a transition from a quadratic to a quartic behavior at an intermediate crossover scale. For suitable model parameters, the tachyonic nature of the region connecting the plateau to the quadratic regime facilitates the formation of oscillons \cite{Piani:2023aof}, being this process largely unaffected by the existence of a quartic regime at small field values if the amplitude of the inflaton field at the beginning of preheating significantly exceeds the aforementioned crossover scale. However, the eventual convergence of the inflaton field amplitude to the quartic regime is expected to strongly impact oscillon stability, ultimately triggering their complete decay and the release of a substantial amount of scalar radiation.

To evaluate the overall lifetime of oscillons in this scenario, we employ a combination of fully-fledged 3+1 lattice simulations and detailed 1+1 radial simulations. Specifically, we carry out a comprehensive 3+1 numerical study of the formation and early evolution of oscillons until they attain a quasi-spherical shape. From this point on, and in order to address the limitations of our lattice resolution as the Universe expands, we follow the approach in \cite{Antusch:2019qrr}, extracting the radial profiles of oscillons and subsequently evolving them using 1+1 radial simulations. This analysis uncovers a distinctive feature: unlike most examples discussed in the literature, oscillons in these scalar-tensor theories exhibit relatively short lifetimes, as a consequence of self-interactions becoming significant at small field values. This unique behavior enables them to emit a substantial amount of radiation, efficiently driving the Universe into a radiation-dominated era without the need for additional fields or interactions.

 This paper is structured as follows. In Section \ref{sec:model}, we introduce the model under consideration, reformulate it into a framework conducive to analyzing inflation and preheating, and examine the potential's behavior across all relevant scales, highlighting the key features of the preheating phase. In Section \ref{sec:Oscillons}, we detail the simulation of oscillon formation and evolution in both 3+1 and 1+1 dimensions, determining the onset of radiation domination, the precise number of $e$-folds of inflation entering CMB observables and the expected spectrum of stochastic gravitational waves associated with oscillon formation. Finally, in Section \ref{sec:conclusions}, we summarize our findings and discuss the implications and scope of the results.
In what follows, we employ a mostly plus metric $(-,+,+,+)$ and we set the Planck mass $M_P=1$. 

%%%%%%%%%%%%%%%%%%%%%%%%%%%%%%%%%%%%%%%%%%%%%%%%%%%%%%%%%%%%%%%%%%%%%%%%
\section{Inflation in Einstein-Cartan scalar-tensor theories} \label{sec:model}
%%%%%%%%%%%%%%%%%%%%%%%%%%%%%%%%%%%%%%%%%%%%%%%%%%%%%%%%%%%%%%%%%%%%%%%%

Compared to General Relativity, EC gravity offers several compelling advantages. First, it can be viewed as a gauge theory of the Poincaré group, aligning gravitational interactions with other fundamental forces in Nature \cite{Utiyama:1956sy,Kibble:1961ba}. Second, the independent treatment of metric and connection in this framework eliminates the need for specific boundary terms to derive the Einstein’s equations of motion \cite{Ferraris1982}. Moreover, the spin connection in EC gravity naturally couples to fermions, providing a powerful arena for incorporating the Standard Model in the presence of gravity \cite{Shaposhnikov:2020aen,Karananas:2021zkl}. Additionally, when non-minimal couplings are introduced, EC gravity offers a versatile foundation for scalar-tensor theories, accommodating both the metric and Palatini formulations as limiting cases.

In this work, we focus on Jordan-frame EC formulations of scalar-tensor theories involving polynomial operators of dimension $d\leq 4$ and no more than two field derivatives. The first condition ensures that non-renormalizable effects arise only from gravitational interactions, thereby preserving the consistency with general relativity. The second condition precludes the introduction of additional propagating degrees of freedom beyond the inflaton field and the two massless graviton polarizations.~\footnote{Note that this condition is sufficient but not necessary. Indeed, as shown in \cite{Karananas:2021zkl,Barker:2024dhb}, it is possible to build models involving higher orders of curvature but not featuring additional
propagating degrees of freedom.} Under these premises, the family of local, Lorentz-invariant actions to be considered in this paper takes the form
\begin{equation}\label{S-EH}
  S[g, \varphi, T^2,\partial T]  =\int d^4x\sqrt{- g} \left[ \frac{ f(\varphi)}{2}
  R-\frac{1}{2}  g^{\mu\nu}\partial_\mu \varphi \partial_\nu \varphi- U(\varphi)\right]+S_T[T^2,\partial T]\,,
\end{equation} 
with 
\begin{equation}
    \label{eq:function-HINY0}
  f(\varphi)=1+\xi \varphi^2 \,,
    \hspace{15mm} 
      U(\varphi)=\frac{\lambda}{4}   \varphi^4\,, 
\end{equation}
$\xi$ and $\lambda$ dimensionless couplings and $S_T$ including all possible operators associated with torsion, namely \cite{Langvik:2020nrs,Shaposhnikov:2020gts,Karananas:2021zkl} 
\begin{equation}
\begin{aligned}
\label{eq:action_torsion}
S_{\rm T} &= \int d^4 x \,\sqrt{-g}\Bigg[ v^\mu \partial_\mu Z^v + a^\mu \partial_\mu Z^a \\
&+ \frac{1}{2}\Big(G_{vv} v_\mu v^\mu  + 2G_{va}v_\mu a^\mu + G_{aa}	a_\mu a^\mu 
 +G_{\tau\tau}\tau_{\alpha\beta\gamma} \tau^{\alpha\beta\gamma}+ \tilde{G}_{\tau\tau} \epsilon^{\mu \nu \rho \sigma} \tau_{\lambda\mu\nu} \tau^\lambda_{~\rho\sigma}\Big) \Bigg]\,,
\end{aligned}
\end{equation}
with 
\begin{equation}
Z^{v/a}=\zeta^{v/a}_\varphi \varphi^2 \,,  \hspace{15mm} 
G_{ij}= c_{ij} \left(1+\xi_{ij}\varphi^2\right) \,,
\end{equation} 
no summation on repeated $i,j$ indices and $\zeta^{v/a}_\varphi$, $c_{ij}$ and $\xi_{ij}$ constants. Here
\begin{equation}
v_\mu = T^\nu_{~\mu\nu} \,, \hspace{10mm} 
a_\mu = \epsilon_{\mu\nu\rho\sigma}T^{\nu\rho\sigma} \,,
\hspace{10mm}
\tau_{\mu\nu\rho} =\frac 2 3 \left( T_{\mu\nu\rho} -v_{[\nu} g_{\rho]\mu} - T_{[\nu\rho]\mu} \right) \ ,
\end{equation}
stand respectively for the vector, pseudo-vector and reduced tensor irreducible components of the torsion tensor,
\begin{equation}
\label{eq:tors_irreps}
T_{\mu\nu\rho} = e_{\mu A} T^A_{\nu\rho}= \frac23 v_{[\nu}g_{\rho]\mu} - \frac16 a^\sigma \epsilon_{\mu\nu\rho\sigma} +\tau_{\mu\nu\rho} \,,
\end{equation}
with the square brackets standing for anti-symmetrization in the corresponding indices. Note that, motivated by the analogy with the Standard Model Higgs field in the unitary gauge, we have assumed our otherwise generic scalar field $\varphi$ to exhibit a $\mathbb{Z}_2$-symmetry and be governed by a purely quartic, scale-invariant potential. All our results continue to hold, however, even in the presence of a quadratic mass term, provided the associated mass parameter remains significantly smaller than the crossover scale.

In order to get a more intuitive picture, we can remove the non-dynamical torsion terms by solving the algebraic equation of motion for the irreducible torsion components
\begin{equation}
    \frac{\delta S}{\delta v_\mu}=0, \quad \frac{\delta S}{\delta a_\mu}=0, \quad \frac{\delta S}{\delta \tau_{\mu \nu \lambda}}=0\,,
\end{equation}
and plug the solution back into the action. This leaves us with a completely equivalent metric theory, where only curvature is present, and the kinetic term of the inflaton is modified, leading to the torsion-free action
\begin{equation}
    \label{eq:Jordan-torsion-free}
      S =\int d^4 x \sqrt{-g} \left[ \frac{f(\varphi)}{2}R - \frac{1}{2}\left(1+\Delta K(\varphi)\right)\partial_\mu \varphi \partial_\nu \varphi  - V(\varphi)\right] \,,
\end{equation}
with the modified kinetic term being
\begin{equation}
    \label{eq:Kj}
     \Delta K(\varphi)=\frac{c_1 \varphi^2+c_2\varphi^4}{1+c_3 \varphi^2+c_4  \varphi^4}
\end{equation}
where the coefficients $c_i$ are expressed in terms of the previously introduced parameters $\{\zeta_\varphi^{v/a},c_{ij},\xi_{ij}\}$ (see~\cite{Rigouzzo:2022yan} for details).
For the purpose of studying the post-inflationary evolution, it is more convenient, however,  to perform a Weyl rescaling of the metric to the Einstein frame ${g}_{\mu\nu}\to f^{-1}(\varphi) g_{\mu\nu}$, which leads to the action
\begin{equation}
    S =\int d^4 x \sqrt{-g} \left[ \frac{R}{2} - \frac{1}{2} K(\varphi)\partial_\mu \varphi \partial_\nu \varphi  - V(\varphi)\right] \,,
\end{equation}
with 
\begin{equation}
\label{eq:K-V}
    K(\varphi)=\frac{1+ c(\varphi) \, \varphi^2}{(1+\xi \varphi^2)^2}\,, \hspace{15mm} V(\varphi)=\frac{U(\varphi)}{(1+\xi \varphi^2)^2}\,,
\end{equation}
and 
\begin{equation}
    c(\varphi) = \xi + 6\xi^2 + 4f(\varphi) \frac{G_{aa}(\zeta^{v}_\varphi)^2 + G_{vv}(\zeta^{a}_\varphi)^2 - G_{va}\zeta^{v}_\varphi\zeta^{a}_\varphi}{G_{vv}G_{aa}-G_{va}^2}
\end{equation}
a function of $\varphi$ determined by all the dimensionless parameters\footnote{Notice that $G_{\tau\tau}$ and $\tilde{G}_{\tau\tau}$ do not appear in the Einstein-frame kinetic term, as the equation of motion for the tensor component of the torsion tensor yeld $\tau_{\mu\nu\lambda}=0$~\cite{Gialamas:2024iyu}.} in \eqref{S-EH}. Written in this form, the effects of torsion are completely encoded within a specific set of higher-dimensional operators in the scalar sector of the theory.  In essence, the presence of torsion in the geometry naturally imposes well-defined selection rules, constraining the form of the numerous non-renormalizable operators that could be otherwise incorporated within a standard effective field theory framework \cite{Shaposhnikov:2020gts,Karananas:2021zkl,Rigouzzo:2022yan,Rigouzzo:2023sbb}.
\newline
The non-canonical kinetic term in Eq.~\eqref{eq:K-V} can be made canonical by performing an additional field redefinition
\begin{equation}
    \label{eq:generic-canonical}
    \frac{d \phi}{d \varphi}=\sqrt{K(\varphi)}\,.
\end{equation}
Although this transformation can be rather involved for generic forms of $K(\varphi)$, we will be mainly interested in EC scenarios leading to field-independent $c$ values.
%%%%%%%%%%%%%%%%%%%%%%%%%%%%%%%%%%%%%%%%%%%%%%%%%%%%%%%%%%%%%%
\begin{figure*}
    \begin{center}
        \includegraphics[width=0.47\textwidth]{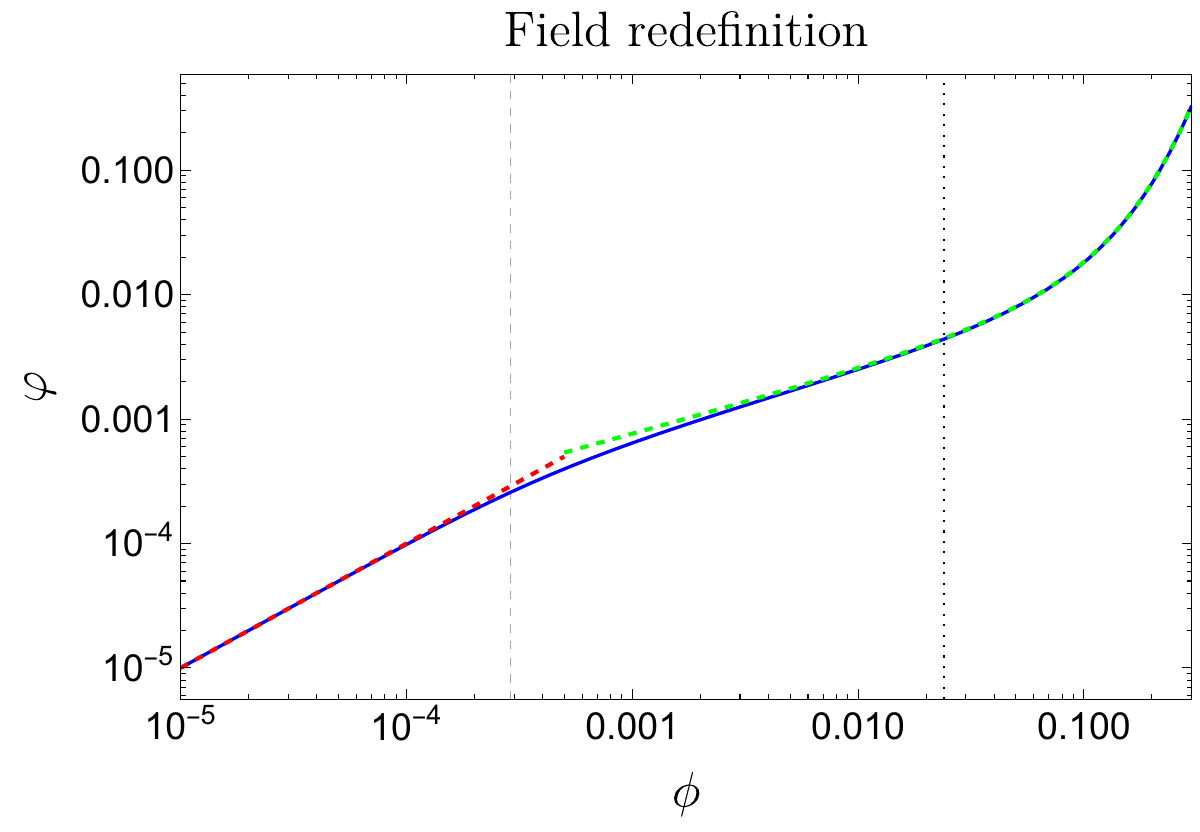} \,\,
                \includegraphics[width=0.47\textwidth]{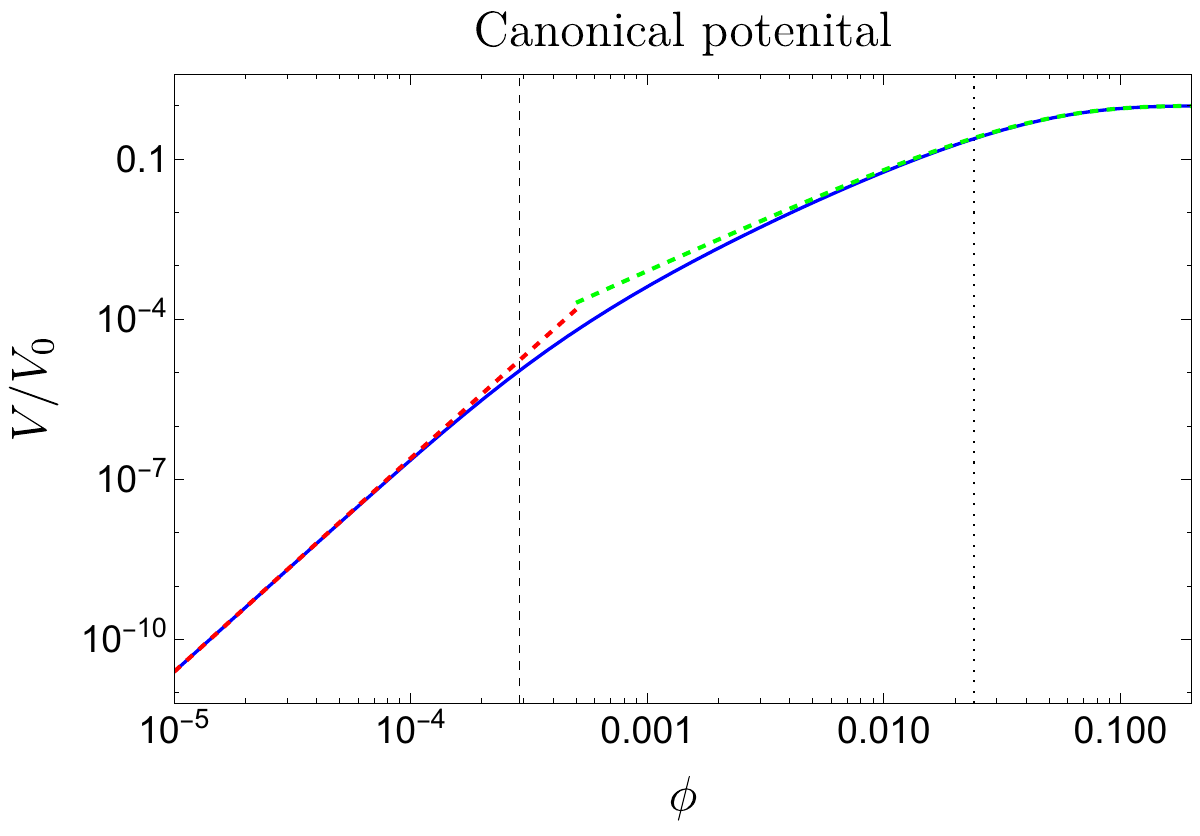}
    \end{center}
    \caption{Non-canonical inflaton field $\varphi$ (left) and canonical Einstein-frame potential (right) as functions of the canonical field $\phi$. The full numerical solutions for a benchmark scenario with $\lambda = 0.001$, $\xi = 5 \cdot 10^4$, and $c = 1.21 \cdot 10^7$ are shown in blue. The small- and large-field approximations, given by Eqs.~\eqref{eq:canonical-field-potential0} and \eqref{eq:canonical-field-potential}, are displayed in red and green, respectively. These approximations are observed to be accurate within their respective ranges of validity. The dashed and dotted vertical lines represent the crossover scale $\phi_c$ and the inflection point $\phi_i$, respectively.
}
    \label{fig:FieldPotential}
\end{figure*}
%%%%%%%%%%%%%%%%%%%%%%%%%%%%%%%%%%%%%%%%%%%%%%%%%%%%%%%%%%%%%%
In this specific yet versatile case,~\footnote{The constant-$c$ condition allows still to describe a plethora of scenarios within the EC scalar-tensor multiverse \cite{Piani:2022gon}. For instance, the choice 
$c_{vv} = -2/3\,, \hspace{2mm} c_{va} =0\,, \hspace{2mm}   c_{aa}=1/24\,, \hspace{2mm} \xi_{vv}=\xi_{aa}=-\zeta^{v}_h=\xi\,, 
\hspace{2mm} \xi_{va}=0\,, \hspace{2mm}
\zeta^{a}_h=\xi_\eta/4,\hspace{1mm}$
leads effectively to a parameter $c=\xi + 6\xi^2_\eta$, with $\xi_\eta$ a non-minimal coupling to be determined from observations. In this limit, the general equation \eqref{eq:action_torsion} reduces to a simple interaction term involving the seminal Nieh-Yan topological invariant~\cite{Nieh:1981ww,Nieh:2008btw}
\begin{equation}
    \label{eq:Nieh-Yan}
    S_{\rm T} =- \frac{1}{4}\int d^4x\:\xi_\eta h^2\partial_{\mu}\left(\sqrt{-g}\epsilon^{\mu\nu\rho\sigma}T_{\nu\rho\sigma}\right)~,
\end{equation} 
with $\epsilon^{\mu\nu\rho\sigma}$ the totally anti-symmetric tensor ($\epsilon_{0123}=1$). 
Alternatively, a choice $\xi_{vv}=\xi_{aa}=\xi$, $c_{av}=0$ translates also into a constant value $c= \xi+  6\xi ^2+ 4\left( \zeta_{ha}^2/c_{aa}+ \zeta_{hv}^2/c_{vv}\right)
$,
allowing to recover the main phenomenology described in this paper for any set of parameters $\xi$, $\zeta_{ha}$, $\zeta_{hv}$, $c_{aa}$, $c_{vv}$ leading numerically to the considered $c$ values. 
} the integral above can be performed analytically, yielding a function $\phi = \phi (\varphi)$ that is generally not invertible for arbitrary values of $\varphi$. However, it can be inverted in both the small and large field approximations, enabling further analytical treatment. In particular, we get 
\begin{equation}
    \label{eq:canonical-field-potential0}
    \varphi\simeq \begin{cases} 
\phi &\hspace{3mm} \textrm{for} \hspace{3mm} \phi<\phi_c~\,,    \\
\xi^{-1/2}\left[\exp\left({\frac{2 \xi \phi}{\sqrt{c}}}\right)-1\right]^{1/2} &\hspace{3mm} \textrm{for} \hspace{3mm} \phi\gg \phi_c~\,, 
   \end{cases}
\end{equation}
with $\phi_c\equiv 1/\sqrt{c}$ a specific crossover scale. The corresponding Einstein-frame potential 
\begin{equation}
    \label{eq:canonical-field-potential}
    V (\phi) \simeq \begin{cases} 
\dfrac{\lambda}{4}\phi^4 &  \hspace{3mm} \textrm{for} \hspace{3mm} \phi<\phi_c~\,,  \\
\dfrac{\lambda}{4 \xi^2}\left[1-\exp\left(- \dfrac{2\xi|\phi|}{\sqrt{c}}\right) \right]^{2}& \hspace{3mm} \textrm{for} \hspace{3mm} \phi \gg\phi_c\,,
   \end{cases}
\end{equation}
exhibits the original quartic behavior at small field values, transitions to a quadratic form with associated mass scale \footnote{The limit $\phi_c\to 0$ in this expression extrapolates the intermediate quadratic behavior of the potential to the origin, effectively neglecting the quartic shape of the potential at $\phi<\phi_c$.} 
\begin{equation}\label{eq:rescaling}
    M^2\equiv V_{,\phi\phi}(0)\vert_{\phi_c\rightarrow0}=\frac{2 \lambda }{c} 
\end{equation} 
at intermediate ones, and approaches a plateau $\lambda/(4\xi^2)$ for field values beyond the inflection point 
\begin{equation}
    \label{eq:inflaction-point}
    \phi_i\equiv \frac{\sqrt{c}\ln 2}{2 \xi}\,.
\end{equation}
For illustration purposes, we display in Fig.~\ref{fig:FieldPotential} the non-canonical field amplitude and the corresponding Einstein-frame potential for a specific set of model parameters, with the crossover scale $\phi_c$ and the inflection point $\phi_i$ explicitly highlighted \footnote{Some of the features of the potential obtained in the constant $c$ scenario can also be inferred from the Jordan-frame action in Eq.~\eqref{eq:Jordan-torsion-free}. In fact, at large $\varphi\gg 1/\sqrt{\xi}$ the theory becomes approximately scale invariant, which corresponds to a plateau in the Einstein frame, while at small field values, where $f(\varphi)\rightarrow1$ and $\Delta K\propto\varphi^2/f(\varphi)\rightarrow 0$, the two frames effectively coincide and one is left with the standard quartic potential. However, the quadratic part connecting these two regime emerges from a non-trivial interplay of $f(\varphi)$ and $\Delta K(\varphi)$. Of course, as the whole procedure is just a change of variables, the results that will follow could be derived directly in the Jordan frame, although it would entail more involved equations, thus not being ideal for numerical approaches. }. 

%%%%%%%%%%%%%%%%%%%%%%%%%%%%%%%%%%%%%%%%%%%%%%%%%%%%%%%%%%%%%%%%%%%%%%%%%%%%%%%%
\subsection{Inflationary dynamics}
%%%%%%%%%%%%%%%%%%%%%%%%%%%%%%%%%%%%%%%%%%%%%%%%%%%%%%%%%%%%%%%%%%%%%%%%%%%%%%%%

Once the canonical Einstein-frame potential has been established, the standard inflationary analysis can be carried out, starting with the calculation of the slow-roll parameters,
\begin{equation}
    \label{eq:slow-roll}
    \epsilon(\phi)=\frac{1}{2}\left( \frac{V_{,\phi}}{V}\right)^2=\frac{8 \xi^2}{c}\left[1- \exp\left(\frac{2 \xi |\phi|}{\sqrt{c}} \right)\right]^2\,, \hspace{5mm} \eta(\phi)=-\frac{8\xi^2}{c}\frac{-2+ \exp\left(\frac{2 \xi |\phi|}{\sqrt{c}} \right)}{\left(1- \exp\left(\frac{2 \xi |\phi|}{\sqrt{c}} \right)\right)^2}\,,
    \end{equation}
the value of the inflaton field at the end of inflation ($\epsilon=1$), 
\begin{equation}
    \label{eq:end-of-inflation}
    |\phi_{\rm end }|= \frac{\sqrt{c} \ln\left(1+2^{3/2} \xi \,c^{-1/2} \right)}{2 \xi}\,,
\end{equation}
and the number of $e$-folds before its end,
\begin{equation}
\label{eq:e-folds0}
    \mathcal{N}=\int_{\phi_{\rm end}}^\phi \frac{d \phi'}{\sqrt{2 \epsilon(\phi')}}\simeq \frac{c}{8 \, \xi^2}e^{2 |\phi| \frac{\xi}{\sqrt{c}}}\quad \longrightarrow \quad |\phi(\mathcal{N})|\simeq \frac{\sqrt{c}}{2 \xi}\ln\left( \frac{8 \xi \mathcal{N}}{c}\right)\,.
\end{equation}
Using these relations, the key primordial CMB observables—namely, the amplitude of the power spectrum of density perturbations $A_s$, the spectral tilt $n_s$ and the tensor-to-scalar ratio $r$—evaluated at the pivot scale $k_* = 0.05 \, \mathrm{Mpc}^{-1}$, can be expressed as
\begin{equation}
    \label{eq:observables}
    A_s=\frac{V_*}{24 \pi^2 \epsilon_*}=\frac{\lambda \mathcal{N}_*^2}{12 \pi^2 c}\,,\hspace{5mm}n_s=1+2\eta_*-6\epsilon_*=1-\frac{2}{\mathcal{N_*}}\,,\hspace{5mm} r=16\epsilon_*=\frac{2c}{\xi^2\mathcal{N}_*^2}\,.
\end{equation}For $\xi = \sqrt{c/6}$ these predictions reduce to those obtained in metric Higgs inflation scenarios ~\cite{Bezrukov:2007ep,Barbon:2009ya,Bezrukov:2009db,Burgess:2010zq,Bezrukov:2010jz,Giudice:2010ka,Bezrukov:2014bra,Hamada:2014iga,Hamada:2014wna,George:2015nza,Fumagalli:2016lls,Bezrukov:2017dyv} and related scale-invariant settings \cite{Shaposhnikov:2008xb,Garcia-Bellido:2011kqb,Garcia-Bellido:2012npk,Bezrukov:2012hx,Rubio:2014wta,Karananas:2016kyt,Trashorras:2016azl,Casas:2017wjh,Tokareva:2017nng,Casas:2018fum,Shaposhnikov:2018jag,Herrero-Valea:2019hde,Karananas:2020qkp,Rubio:2020zht,Shaposhnikov:2020frq,Karananas:2021gco,Gialamas:2021enw,Piani:2022gon,Belokon:2022pqf,Karananas:2023zgg}, with the spectral tilt and the tensor-to-scalar ratio exhibiting no dependence on the exact value of $\xi$ (for a comprehensive review, see~\cite{Rubio:2018ogq}). In contrast, for $\xi=c$, the above predictions align with those obtained in Palatini Higgs inflation and related extensions~\cite{Bauer:2008zj,Bauer:2010jg,Rasanen:2017ivk,Enckell:2018kkc,Rasanen:2018fom,Rasanen:2018ihz,Gialamas:2019nly,Rubio:2019ypq,Shaposhnikov:2020fdv,Annala:2021zdt,Dux:2022kuk,Poisson:2023tja}, featuring a strong $1/\xi$ suppression on the tensor-to-scalar ratio.

%%%%%%%%%%%%%%%%%%%%%%%%%%%%%%%%%%%%%%%%%%%%%%%%%%%%%%%%%%%%%%%%%%%%%%%%
\subsection{Post-inflationary phenomenology}\label{sec:Pheno}
%%%%%%%%%%%%%%%%%%%%%%%%%%%%%%%%%%%%%%%%%%%%%%%%%%%%%%%%%%%%%%%%%%%%%%%%

After the end of inflation, the inflaton field can be effectively described as a homogeneous condensate, with the majority of its energy concentrated in its zero mode. The field's dynamics is governed by the Klein-Gordon equation in a FLRW spacetime,
\begin{equation}
    \label{eq:KG-FLRW}
    \ddot{\phi}+3 H\dot{\phi}-a^{-2}\nabla^2 \phi+V_{,\phi}=0\,, 
\end{equation}
with $a$ the scale factor, $H\equiv\dot{a}/a$ the Hubble rate dictated by the Friedmann equations
\begin{equation}
\label{eq:Friedmann}
    3 H^2 =\frac{1}{2}\dot{\phi}^2+ \frac{1}{2a^2}(\nabla\phi)^2  + V(\phi) \, , \hspace{10mm} \quad \dot{H} = -\frac{1}{2} \dot{\phi}^2-\frac{1}{6a^2} (\nabla\phi)^2 \,,
\end{equation}
and the overdots denoting differentiation with respect to cosmic time. In the absence of particle production, the inflaton field would oscillate around the minimum of its potential with a decaying amplitude $\phi \sim a^{-3(1+\bar{w})}$ dictated by the time-averaged equation of state \cite{Turner:1983he}
\begin{equation}
    \label{eq:eos-monomial}
    \bar{w} \equiv \frac{\langle p\rangle}{\langle \rho\rangle} \simeq \frac{n-1}{n+1}\,,
\end{equation}
with $n=1$ ($\bar{w}=0$) in the quadratic regime ($\phi>\phi_c$) and $n=2$ ($\bar{w}=1/3$) in the quartic one ($\phi<\phi_c$). However, the non-linear preheating dynamics can significantly influence the post-inflationary evolution, even in the absence of direct interactions with other matter fields. While fully capturing these phenomena requires solving the exact equations of motion numerically, the initial stages of particle production can be adequately described by a linearized analysis. To this end, let us decompose the inflaton as a sum of background and fluctuations, $\phi (\vec{x}, t)=\bar{\phi}(t)+\delta\phi (\vec{x},t)$, and examine the evolution of the latter in Fourier space.  At leading order, the Fourier modes  $\delta \phi_{{\bf k}}$ satisfy the equation 
\begin{equation}
    \label{eq:KG-Fourier}
    \ddot{\delta \phi_{\textbf{k}}}+3 H\dot{\delta \phi_{\textbf{k}}} +\left(\textbf{k}^2/a^2 +V_{,\phi\phi} (\bar{\phi})\right)\delta \phi_{\textbf{k}}=0\,,
\end{equation}
with the background field $\bar{\phi}$, and consequently the effective mass $V_{,\phi\phi}$ evolving over time. Depending on the average field amplitude within the potential \eqref{eq:canonical-field-potential}, we can distinguish three distinct regimes:
\begin{itemize}
    \item \textit{\underline{Tachyonic region} } (${ |\phi|>\phi_i}$):   For field values in this range, the second derivative of the potential \eqref{eq:canonical-field-potential} becomes negative, triggering a tachyonic instability in Eq.~\eqref{eq:KG-Fourier} and the exponential growth of long-wavelength fluctuations with $ \textbf{k}^2<a^2|V_{,\phi\phi}|$. The maximum momentum subject to tachyonic amplification is determined by the minimum of the second derivative of the potential, ${\rm min}\{V_{,\phi\phi}\}$, which occurs at a parametrically fixed value
\begin{equation}
    \label{eq:min-k}
     \phi_{\rm tach}=\frac{\sqrt{c} \ln 2}{\xi}\hspace{5mm} \longrightarrow \hspace{5mm} \frac{|\mathbf{k}_{\rm max}|}{a}=\sqrt{|V_{,\phi\phi}(\phi_{\rm tach})|}=2^{-3/2} M \simeq 0.35 M\,.
\end{equation}
\item \textit{\underline{Quadratic region}} (${\phi_c<|\phi|<\phi_i}$): For field values between the inflection point and the crossover scale, the potential \eqref{eq:canonical-field-potential} behaves, at leading order, as a quadratic potential. In this regime, Eq.~\eqref{eq:KG-Fourier} reduces to the equation of motion of a damped harmonic oscillator with effective mass $M$. As a result, particle production is significantly suppressed.
\item \textit{\underline{Quartic region} (${|\phi|<\phi_c}$):} At small field values, the model effectively reduces to a $\phi^4$ theory, resulting in particle production through parametric self-resonance \cite{Lozanov:2016hid,Lozanov:2017hjm}. The efficiency of this heating mechanism is nonetheless significantly weaker than the tachyonic amplification at larger field amplitudes.
\end{itemize}
The actual post-inflationary dynamics of the system will strongly depend on the interplay and timing of these regimes, with the metric and Palatini scenarios studied in Refs.~\cite{Garcia-Bellido:2008ycs,Bezrukov:2008ut,Bezrukov:2014ipa,Rubio:2015zia,Repond:2016sol,DeCross:2015uza,DeCross:2016fdz,DeCross:2016cbs,Ema:2016dny,Sfakianakis:2018lzf} and \cite{Rubio:2019ypq,Dux:2022kuk} respectively as limiting cases. As shown in \cite{Piani:2023aof}, oscillons successfully form if the following two conditions are satisfied:
\begin{enumerate}
    \item There is a sufficient separation between the inflaton amplitude at the onset of preheating $\phi_{\rm end}$ and the inflection point $\phi_i$,
    \item The crossover scale $\phi_c$ is much smaller than the typical oscillation amplitude of the inflaton at the end of inflation.
\end{enumerate}
Similarly to the Palatini scenario studied in \cite{Rubio:2019ypq,Dux:2022kuk}, the first condition allows the field to re-enter the tachyonic region for several semi-oscillations. This leads to multiple consecutive periods of instability where sub-horizon inflaton modes below the maximum unstable momentum $|\textbf{k}|<|\textbf{k}_{\rm max}|$ grow exponentially. Depending on the specific parameter choice, these consecutive instabilities can cause some of the modes to grow and reach a non-linear regime, where their energy becomes comparable to that of the background, eventually causing backreaction and the fragmentation of the condensate. If the second condition is also satisfied, the resulting fragments can cluster and form oscillons, without the quartic self-interaction at small field values being relevant enough to spoil their formation.
Nonetheless, as we will show in what follows, this interaction at the bottom of the potential can play a relevant role once the oscillons have formed, modifying the overall dynamics and lifetime.

%%%%%%%%%%%%%%%%%%%%%%%%%%%%%%%%%%%%%%%%%%%%%%%%%%%%%%%%%%%%%%%%%%%%%%%%
\section{Oscillon dynamics}\label{sec:Oscillons}
%%%%%%%%%%%%%%%%%%%%%%%%%%%%%%%%%%%%%%%%%%%%%%%%%%%%%%%%%%%%%%%%%%%%%%%%

As explained in the introduction, the presence of oscillons can significantly influence the post-inflationary evolution, potentially leading to an extended period of matter domination. To accurately estimate the inflationary observables, it is therefore crucial to determine the time required for oscillons to form, the fraction of the Universe's energy density stored in them, and the duration of their stability before decaying. For concreteness, we will focus in what follows on a specific choice of parameters satisfying the above conditions for oscillon formation, 
\begin{equation}
    \label{eq:param}
    \lambda=0.001\,, \hspace{5mm} \xi=5\cdot10^4 \,,\hspace{5mm} c=1.21\cdot10^7 \ ,
\end{equation}
which coincides with one of the cases considered in~\cite{Piani:2023aof}. For such a purpose, we roughly divide the oscillons' evolution into two stages: their formation and their decay. 
\begin{itemize}
    \item In Section \ref{sec:OscInit} we perform lattice simulations of the post-inflationary dynamics in 3+1 dimensions. These simulations enable us  to capture the formation and early evolution of oscillons, and to extract their radial profiles once they stabilize and achieve a quasi-spherical shape.
    \item In Section \ref{sec:OscLate}, we investigate the late-time evolution of oscillons by evolving each individual oscillon in 1+1 dimensions under the assumption of spherical symmetry and with the radial profiles obtained in the previous 3+1 lattice simulations as initial conditions. This approach allows us to observe the oscillons' decay and parameterize their lifetimes as a function of their initial shapes. Additionally, we calculate the energy transfer rate to radiation, which enables us to characterize the transition from matter  domination to radiation domination.
\end{itemize}
Finally, in Section \ref{sec:results} we discuss how the details of the transition between inflation and radiation domination affect the inflationary observables, as well as the gravitational wave signal produced by the oscillons. 

%%%%%%%%%%%%%%%%%%%%%%%%%%%%%%%%%%%%%%%%%%%%%%%%%%%%%%%%%%%%%%%%%%%%%%%%
\subsection{Formation and early-time evolution} \label{sec:OscInit}
%%%%%%%%%%%%%%%%%%%%%%%%%%%%%%%%%%%%%%%%%%%%%%%%%%%%%%%%%%%%%%%%%%%%%%%%

To characterize the formation and initial evolution of oscillons, we resort to 3+1 classical lattice simulations carried out with the publicly available code \CL \cite{Figueroa:2021yhd,Figueroa:2023xmq}. This program solves self-consistently the Klein-Gordon and scale factor equations \eqref{eq:KG-FLRW}-\eqref{eq:Friedmann} on a regular cubic lattice of side length \(L\), with \(N\) points per dimension and periodic boundary conditions. The initial conditions for our simulations are set at the end of inflation, defining \(t=0\) and \(a(0)=1\). The initial homogeneous components of the inflaton field and its time derivative are determined by solving the homogeneous equations of motion from deep into the slow-roll regime up to the end of inflation. On top of these background components, we superimpose quantum vacuum fluctuations below a specified momentum cutoff scale. Finally, the parameters \(N\) and \(L\) are selected to ensure adequate resolution of the modes excited after inflation. For our benchmark simulation, we set \(N = 288\) and \(L = 15.7 \,M^{-1}\), which enables us to resolve momenta between \(k_{\rm IR} \equiv 2\pi/L = 0.4\,M\) and \(k_{\rm UV} \equiv (\sqrt{3} N / 2) k_{\rm IR} = 100\,M\). 

\begin{figure*}
    \begin{center}
        \includegraphics[width=0.65\textwidth]{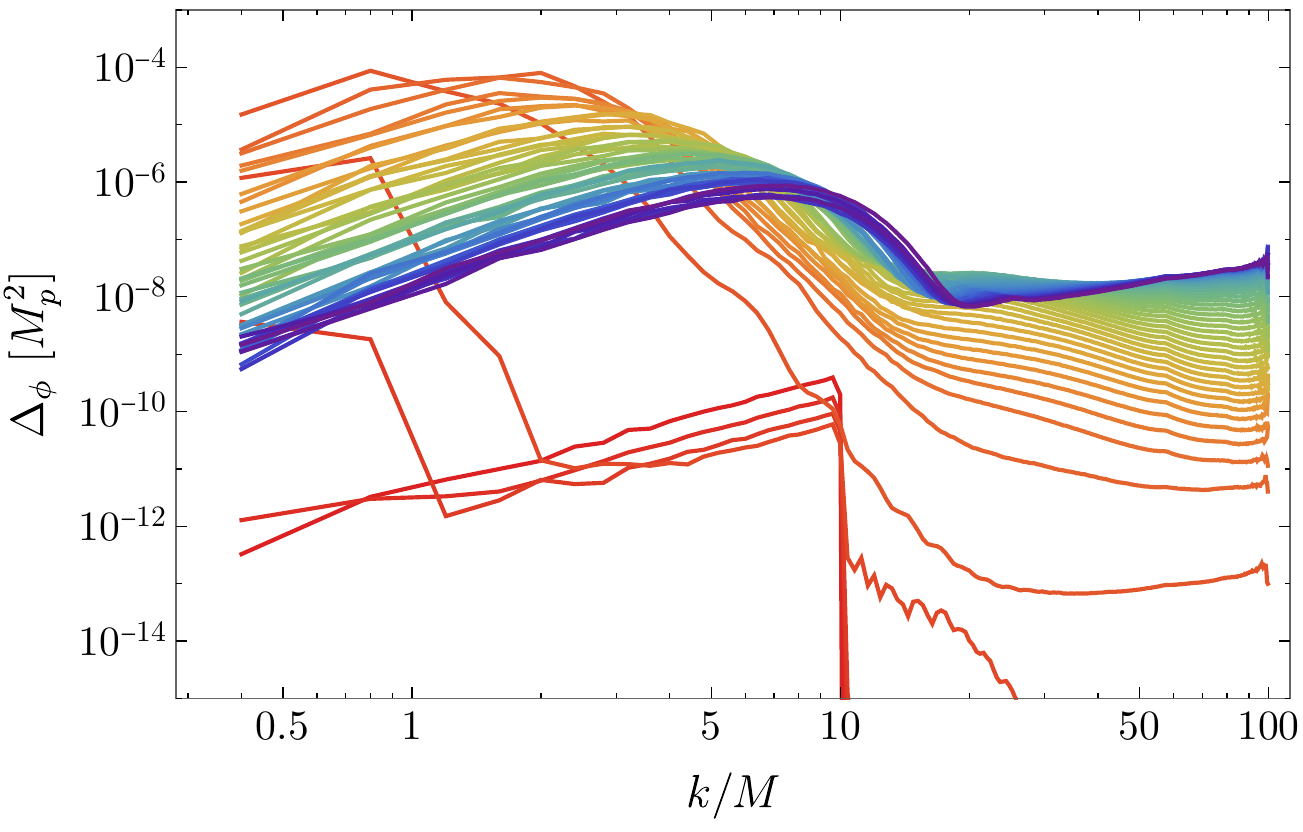}
    \end{center}
    \caption{Scalar field power spectrum $\Delta_{\phi} \equiv k^3 |\phi_k|^2 /(2 \pi^2)$ extracted from the 3+1 lattice simulations between times $M t= 0$ and $M t=1500$, with a time interval $M \Delta t= 30$. Lines go from red (early times) to purple (late times).}
    \label{fig:SpectraLattice}
\end{figure*}

Fig.~\ref{fig:SpectraLattice} shows the resulting evolution of the inflaton's power spectrum. At early times, only momenta within the infrared band \(0 < k < k_{\rm max}\) are excited due to the tachyonic oscillations of the inflaton's homogeneous mode. Around \(M t = \mathcal{O}(10^2)\), the energy stored in the inflaton fluctuations has grown to the point where non-linearities become significant, and modes with increasingly higher momenta start getting  populated. Beyond this point, a prominent peak emerges in the spectrum at the physical momentum \(k/(a M) \approx 0.5\), which corresponds to the characteristic size of the oscillons.

%%%%%%%%%%%%%%%%%%%%%%%%%%%%%%%%%%%%%%%%%%%%%%%%%%%%%%%%%%%%%%%%%%%%%%%%
%%%%%%%%%%%%%%%%%%%%%%%%%%%%%%%%%%%%%%%%%%%%%%%%%%%%%%%%%%%%%%%%%%%%%%%%

\begin{figure*}
    \begin{center}
        \includegraphics[width=0.4\textwidth]{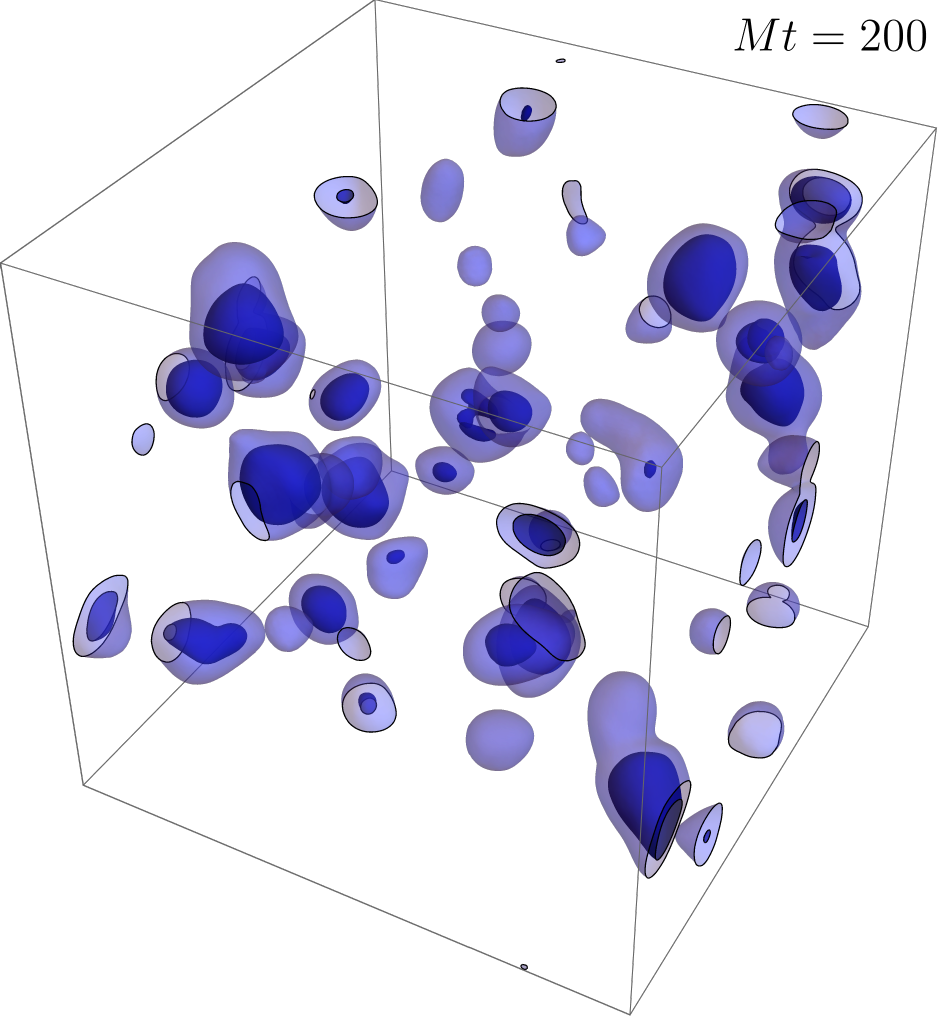}
        \includegraphics[width=0.4\textwidth]{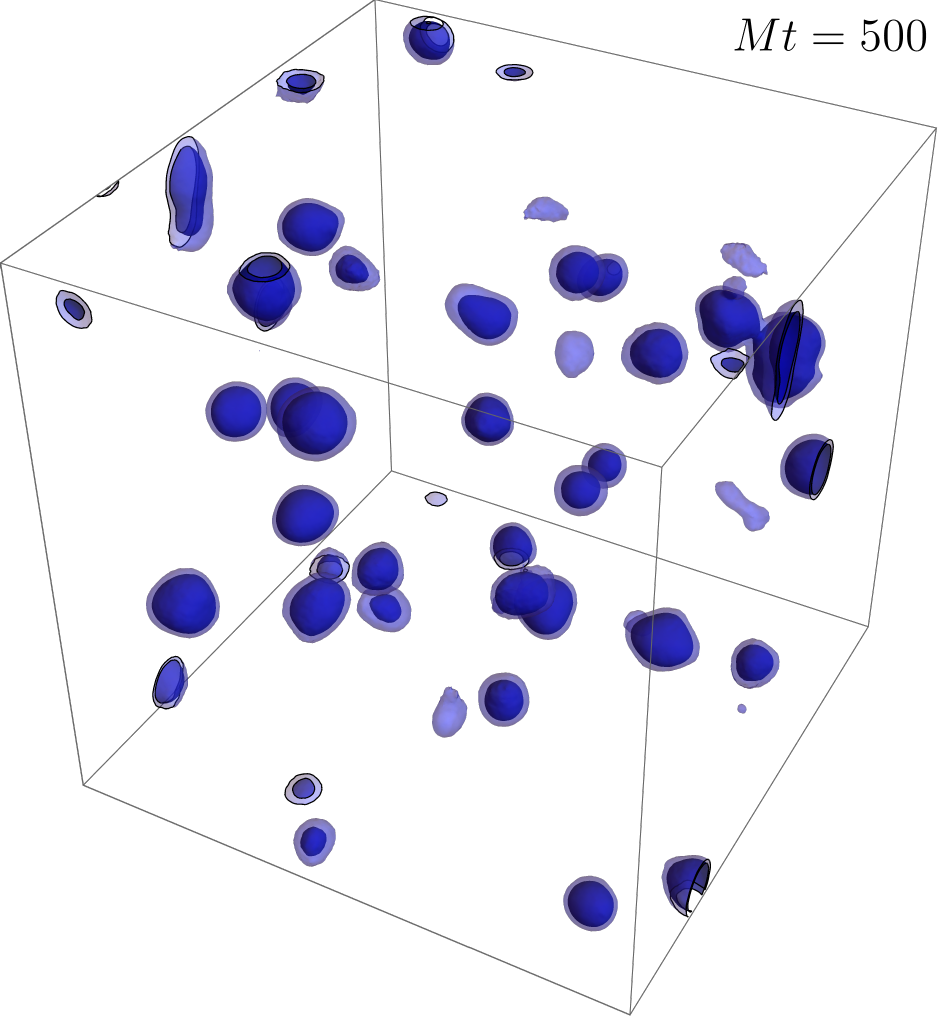} \vspace{0.4cm} \\
        \includegraphics[width=0.4\textwidth]{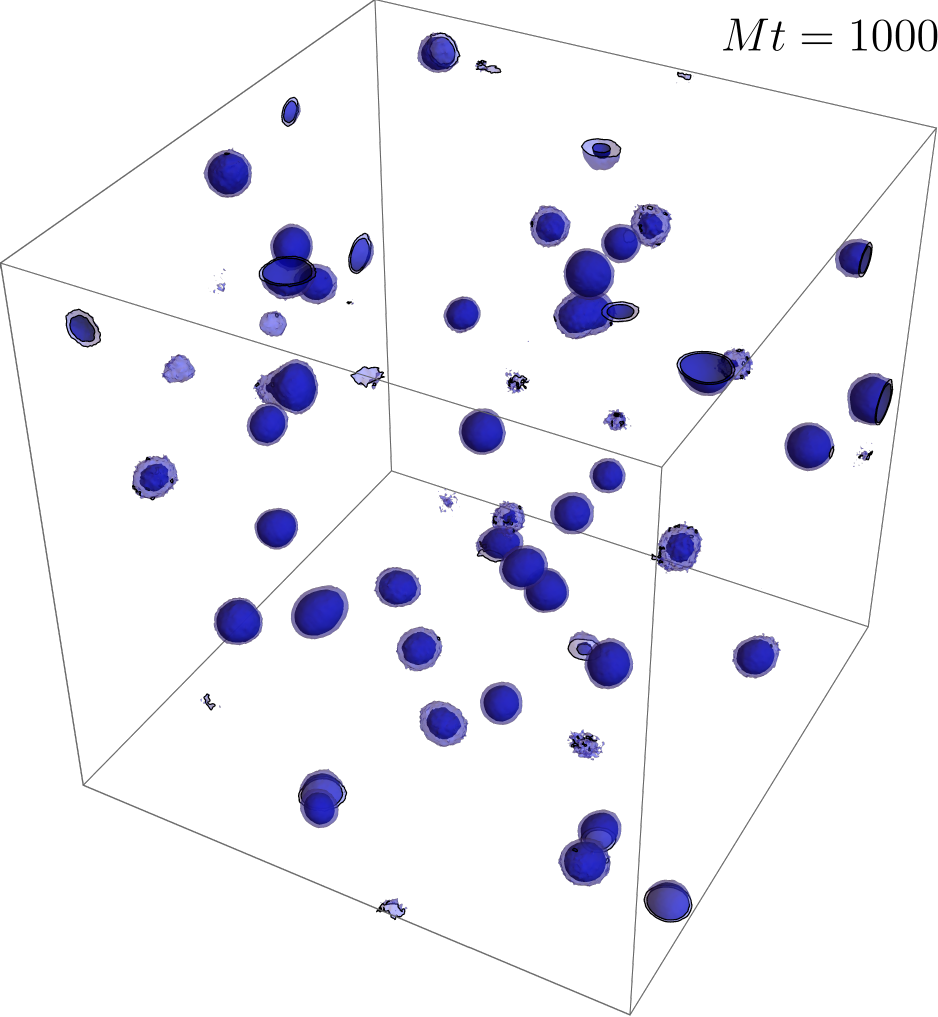}
        \includegraphics[width=0.4\textwidth]{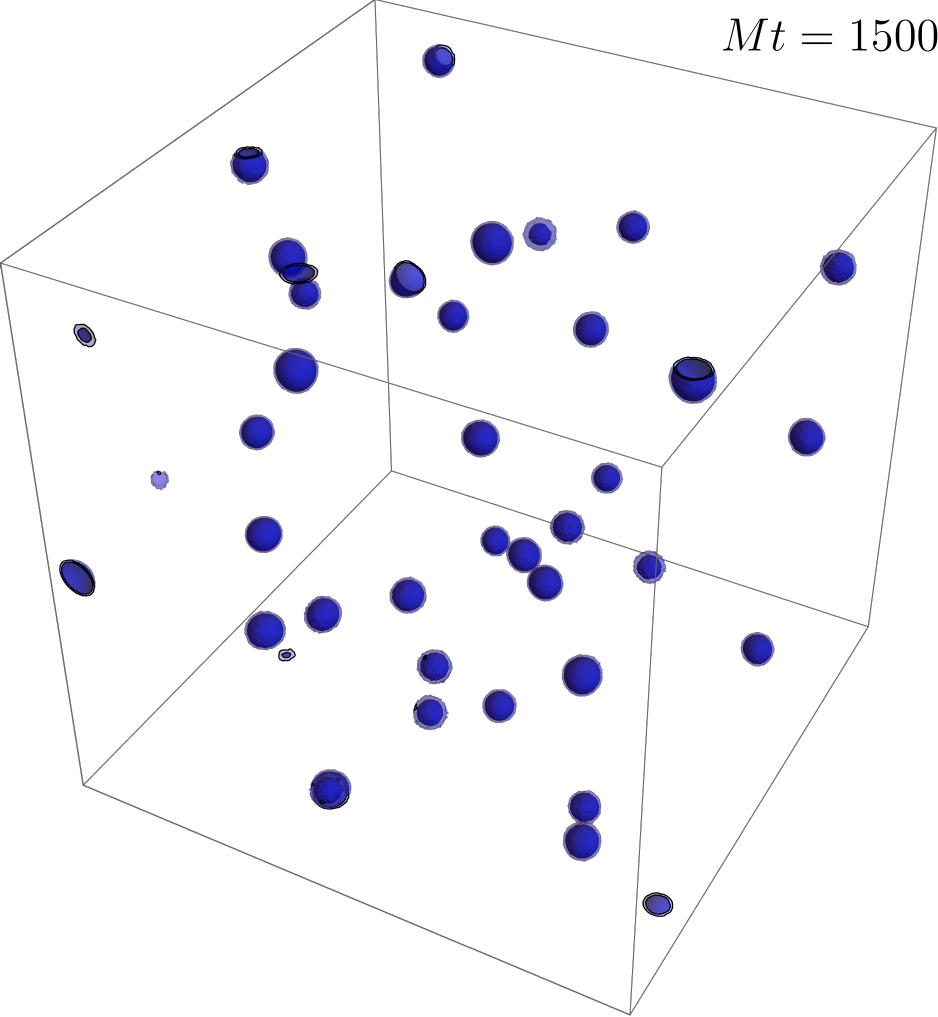} \vspace{0.4cm} \\
        \includegraphics[width=0.4\textwidth]{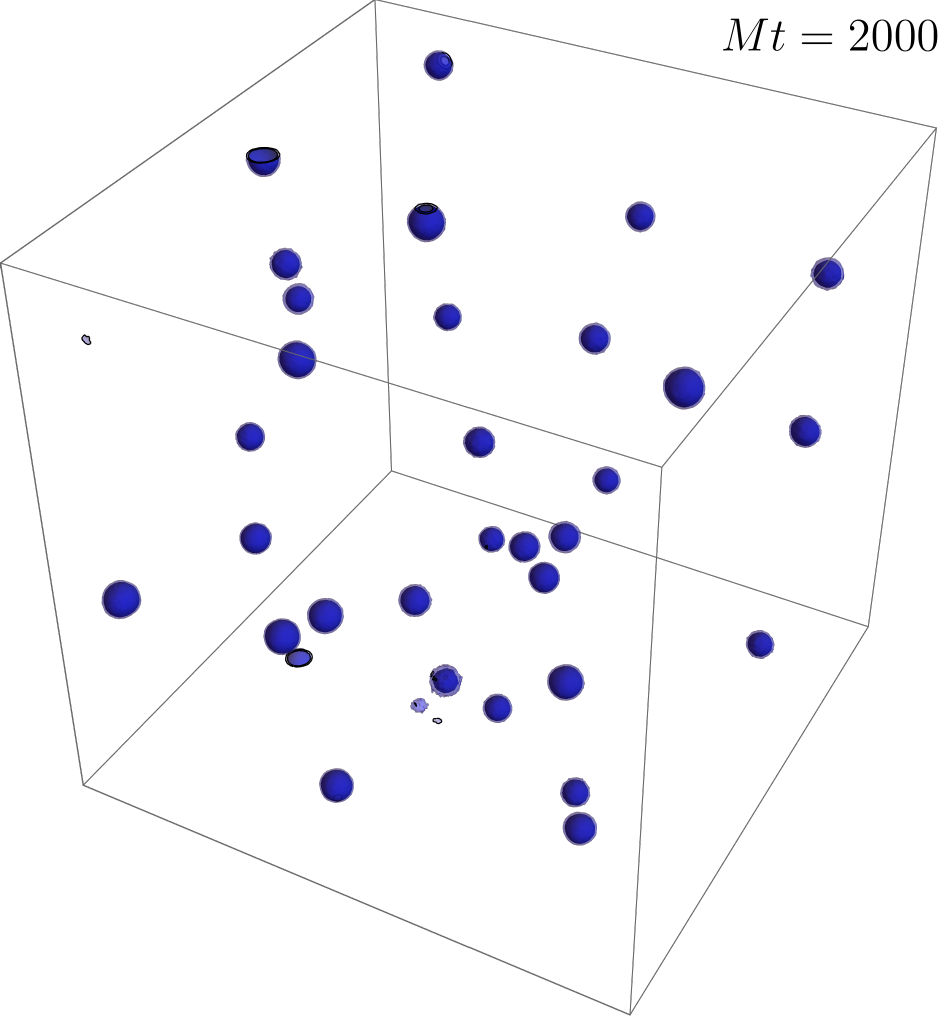}
        \includegraphics[width=0.4\textwidth]{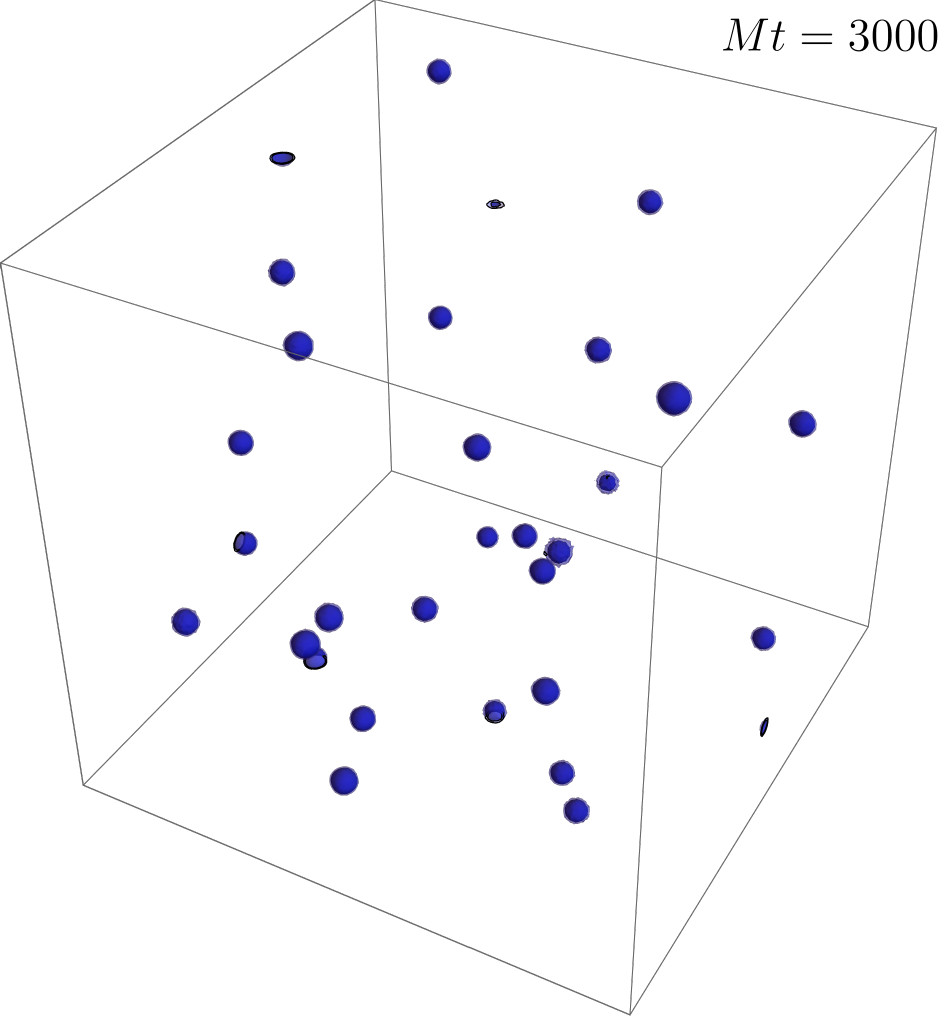} \\
    \end{center}
    \caption{Distribution of the inflaton's energy overdensity extracted from the lattice simulation at times $M t = 200$, $500$, $1000$, $1500$, $2000$, $3000$, corresponding to scale factors $a=4.2$, $7.6$, $12.0$, $15.5$, $18.6$, $24$ respectively. Light/dark blue shades shows the lattice points at which the energy overdensity is $\Delta_{\rho} \equiv \rho / \bar{\rho} = 6, 20$ respectively.}
    \label{fig:OscillonSnapshots}
\end{figure*}

In Fig.~\ref{fig:OscillonSnapshots} we present snapshots of the oscillon distribution at various times. Specifically, we highlight, in different shades of blue, regions where the local energy density exceeds the average by factors of six and twenty times (i.e.~$\Delta_{\rho} (\vec{r}) \equiv \delta \rho (\vec{r}) / \bar{\rho} = 6, 20$). By the time $M t \simeq 200$, oscillons have already formed. However, many spatial regions with high energy overdensities at this time do not correspond to oscillons, but rather to transient structures that decay rapidly. Additionally, at this early stage, the oscillons are far from spherical, making it more appropriate to extract their shapes at later times. As the Universe expands, however, the oscillons' comoving radii decrease, limiting our ability to resolve them at very late times. To balance these two factors, we choose to extract the oscillon shapes at $M t_{\rm ext} = 1500$,  when nearly all transient structures have already decayed and oscillons are both nearly spherical and sufficiently well-resolved at small scales.

%%%%%%%%%%%%%%%%%%%%%%%%%%%%%%%%%%%%%%%%%%%%%%%%%%%%%%%%%%%%%%%%%%%%%%%%
%%%%%%%%%%%%%%%%%%%%%%%%%%%%%%%%%%%%%%%%%%%%%%%%%%%%%%%%%%%%%%%%%%%%%%%%

The oscillons' shapes are extracted as follows. First, we capture five snapshots of the lattice's energy density over a short period of time around $M t \approx 1500$, encompassing approximately 1-2 oscillon oscillations. For each of these snapshots, we identify all connected regions where the local energy overdensity exceeds a threshold $\Delta_{\rho} > 6$. This allows us to identify all oscillons in each snapshot, track their oscillatory evolution, and approximately determine the time at which they reach their maximum amplitude. Next, we fit the shape of each oscillon to the Gaussian function
\begin{equation}
     \phi (r) =  A\,e^{-r^2/R^2} \ . \label{eq:Gaussian-fit}
\end{equation}
where $r$ is the \textit{physical} radial distance between each lattice point and the oscillons' center, defined as the point where the field amplitude attains its absolute maximum $A \equiv \phi(r=0)$, in Planckian units. From this fit, we extract the oscillon's radius $R$.~\footnote{Note that $R$ is simply a fitting parameter used to match the oscillon profiles, and not the physical radial distance at which the overdensity drops below the oscillon-defining value $\Delta_{\rho} = 6$.}

\begin{figure*}
    \begin{center}
        \includegraphics[width=0.47\textwidth]{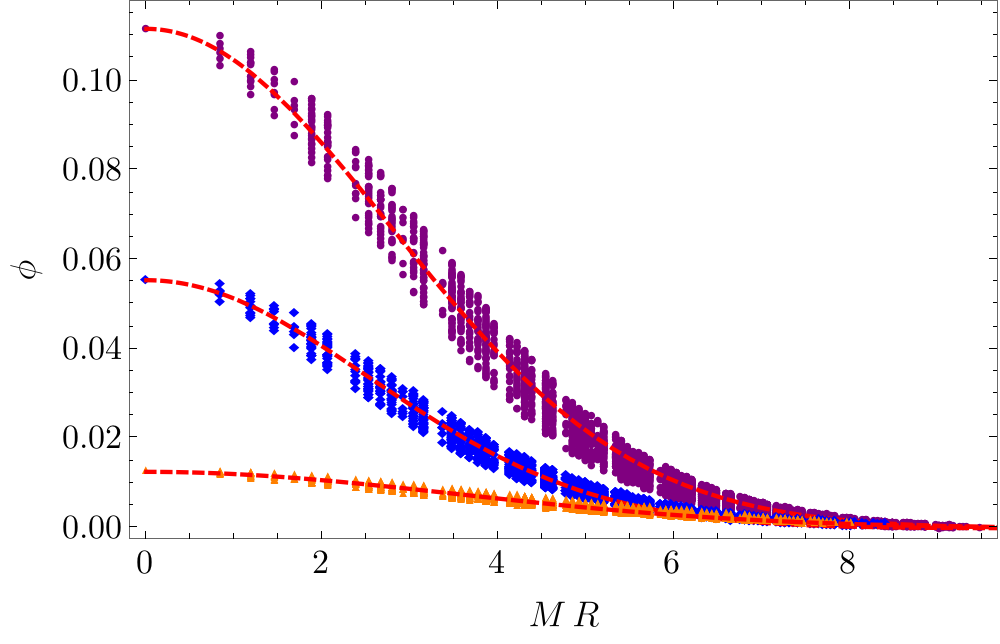} \,\,
                \includegraphics[width=0.47\textwidth]{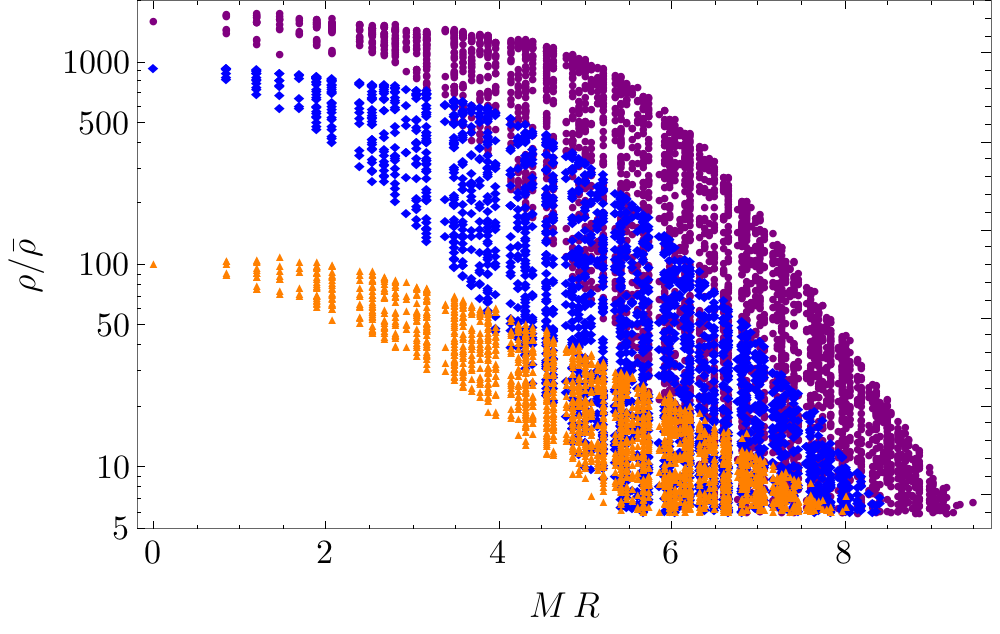}
    \end{center}
    \caption{Left: Radial profiles of three representative oscillons with large, medium, and small amplitudes (purple, blue, and orange, respectively). Each dot represents the field amplitude at an individual lattice site within the oscillon. The red dashed lines indicate the best fits of each oscillon profile to \eqref{eq:ARfit}. Right: Energy overdensity profiles for the same oscillons.}
    \label{fig:ThreeOscillons}
\end{figure*}

\begin{figure*}
\begin{center}
    \includegraphics[width=0.65\textwidth]{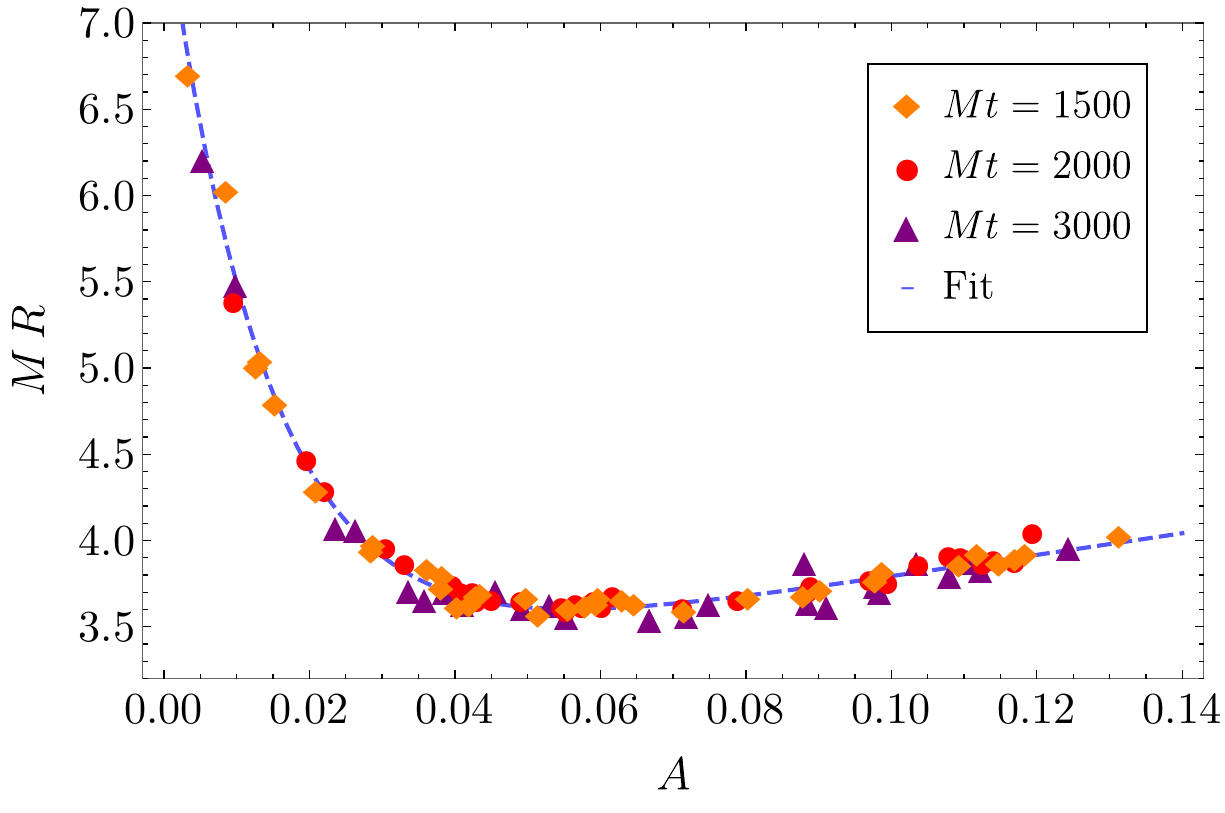}
\end{center}
\caption{Distribution of amplitudes and radii for the oscillons extracted from one 3+1 lattice simulation at times $M t = 1500, 2000, 3000$. The dashed line shows the best fit to the data \eqref{eq:ARfit}.}
\label{fig:AmpRad}
\end{figure*}

We have performed several 3+1 lattice simulations with different initial condition realizations. In each simulation we observe approximately $35$ oscillons, spanning a broad range of amplitudes and radii. For illustration, the left panel of Fig.~\ref{fig:ThreeOscillons} shows the extracted profiles of three representative oscillons with small, medium and large amplitudes ($A=0.015,0.06,0.11$), along with their corresponding Gaussian fits. We see clearly that the different field amplitudes within the oscillon are never too far away from the best fit, indicating that the oscillons are rather spherical at the extraction time. This confirms that the Gaussian fit \eqref{eq:Gaussian-fit} provides an excellent approximation to the oscillons' real shape in all three cases. Note that, despite having different amplitudes, the radii of the oscillons are rather similar, $R \sim 3.5\,M^{-1}$. The corresponding energy overdensity profiles are displayed in the right panel of Fig.~\ref{fig:ThreeOscillons}. As expected, the larger the amplitude of the oscillon, the larger the energy overdensity. The oscillon with the smallest amplitude attains an energy overdensity of $\Delta_{\rho} \sim 10^2$ at the center, while the other two reach energy overdensities of $\Delta_{\rho} \gtrsim 10^3$. In both cases, these overdensities are significantly larger than the threshold $\Delta_{\rho} = 6$ used to extract the oscillons from the lattice.
%%%%%%%%%%%%%%%%%%%%%%%%%%%%%%%%%%%%%%%%%%%%%%%%%%%%%%%%%%%%%%%%%%%%%%%%
%%%%%%%%%%%%%%%%%%%%%%%%%%%%%%%%%%%%%%%%%%%%%%%%%%%%%%%%%%%%%%%%%%%%%%%%
In Fig.~\ref{fig:AmpRad} we show the distribution of amplitudes and radii of all oscillons extracted from the same simulation at times $M t = 1500$, $2000$ and $3000$. We observe that all oscillons are very well distributed along a single curve in the amplitude-radius parameter space, which we have fitted as 
\begin{equation}
M \cdot R= 3.16 + 6.31 A + 4.57 e^{-71.4 \,A} \ . \label{eq:ARfit} 
\end{equation}
We have repeated this analysis for oscillons extracted from 3+1 lattice simulations with different initial fluctuation realizations, confirming that their distribution of amplitudes and radii follows the same behavior. Oscillons lose energy through the emission of field waves and eventually decay completely into radiation. One could a priori extend the 3+1 lattice simulations to later times to observe this phenomenon. However, as seen in Fig.~\ref{fig:SpectraLattice}, the amplitude of the UV tail of the spectrum increases with time, and by $Mt_{\rm ext} = 1500$ it is already only two orders of magnitude below the oscillons' peak amplitude, which decreases over time. At later times, the results become sensitive to the position of the lattice's ultraviolet cutoff, leading to unreliable outcomes.

%%%%%%%%%%%%%%%%%%%%%%%%%%%%%%%%%%%%%%%%%%%%%%%%%%%%%%%%%%%%%%%%%%%%%%%%
%%%%%%%%%%%%%%%%%%%%%%%%%%%%%%%%%%%%%%%%%%%%%%%%%%%%%%%%%%%%%%%%%%%%%%%%

%%%%%%%%%%%%%%%%%%%%%%%%%%%%%%%%%%%%%%%%%%%%%%%%%%%%%%%%%%%%%%%%%%%%%%%%
\subsection{Decay and onset of radiation domination} \label{sec:OscLate}
%%%%%%%%%%%%%%%%%%%%%%%%%%%%%%%%%%%%%%%%%%%%%%%%%%%%%%%%%%%%%%%%%%%%%%%%

To overcome the limitation imposed by simulating in an expanding box, we have simulated the later evolution and decay of each individual oscillon in a 1+1-dimensional regular lattice under the assumption of spherical symmetry $\phi \equiv \phi(r,t)$. More specifically, for each oscillon we solve a discretized version of the radial Klein-Gordon equation
\begin{equation} 
 \frac{\partial^2\phi}{\partial t^2}   - \left( \frac{\partial^2\phi}{\partial r^2} + \frac{2}{r}\frac{\partial\phi}{\partial r} \right) + V_{,\phi}  = 0 \ , \hspace{0.5cm} \phi (r, t=t_{\rm ext}) = A\,  e^{-r^2/R^2} \,,\hspace{5mm}\dot{\phi}(r,t=t_{\rm ext})=0 \,,  \label{eq:rad-eom1}
\end{equation}
where for each oscillon the amplitude $A$ and radius  $R$ are extracted from the 3+1 lattice simulations at the extraction time $Mt_{\rm ext} = 1500$. The center of the oscillon is positioned at the leftmost node of the lattice, and as the oscillon oscillates, the emitted field waves eventually reach the right boundary. In order to prevent unwanted interference effects from reflected waves, we truncate the inflaton amplitude and its time-derivative at intermediate radial distances each time this occurs, ensuring that $\phi= 0$ at the right boundary. Technical details of this setup are provided in Appendix \ref{sec:rad-sim}.

\begin{figure*}
    \begin{center}
    \includegraphics[width=0.47\textwidth]{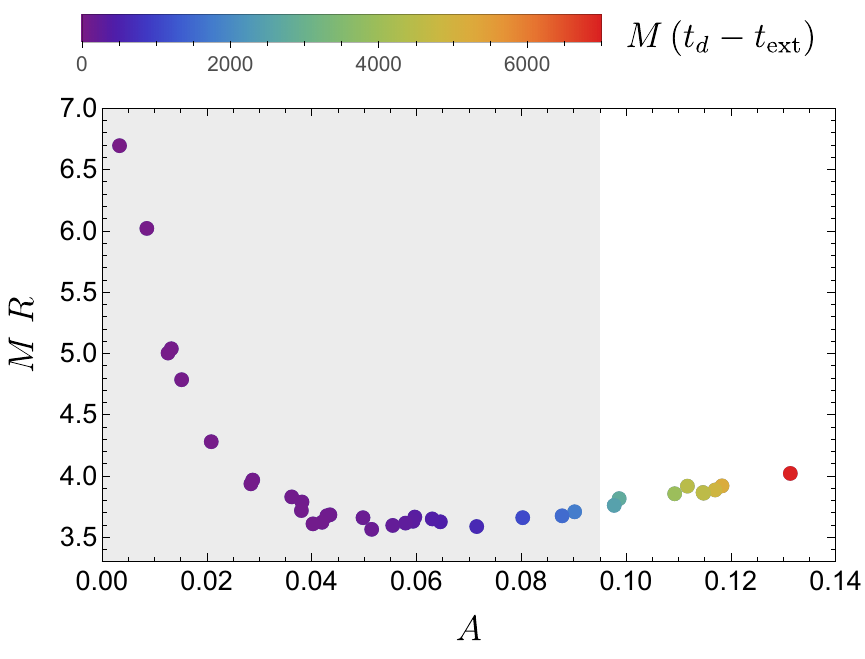}
    \includegraphics[width=0.47\textwidth]{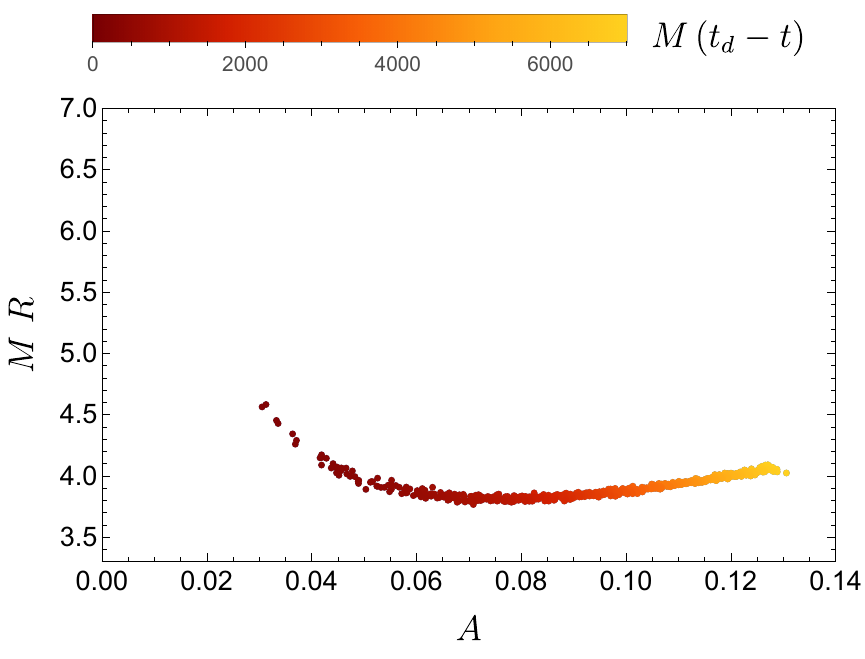} 
         
    \end{center}
    \caption{Left: Lifetime of the oscillons as extracted from the radial simulation. The profile of each oscillon is evolved with a 1+1 radial simulation to find the time $t_d$ at which the energy of each oscillon drops below its value at extraction time. The shadowed region shows oscillons whose summed initial energy is 40\% of the total energy stored in oscillons, in order to highlight the fact that although subdominant in number, oscillons with the larger amplitude dominate the energy contribution. Right: Time evolution of the radial profile of the most energetic oscillon. We take the most energetic oscillon extracted from the 3+1 lattice simulation, and we extract its radial profile coming from the 1+1 radial simulation each time the kinetic energy attains a local minimum. We then perform a Gaussian fit on such profile, thus obtaining the time evolution for A and R. The coloring represents the time left before the oscillon decays.}
    \label{fig:energy-lifetime}
\end{figure*}

These simulations allow us to track the evolution of each oscillon's profile from the initial time $t_{\rm ext}$ until their decay times $t_{\rm dec}$, defined as the moment when the energy of an oscillon drops below $0.1\%$ of its value at the extraction time $E_{\rm ext}^i$. As shown in Fig.~\ref{fig:energy-lifetime}, there is a clear correlation between an oscillon's central amplitude and its lifetime. This is due to the fact that, by the extraction time, the inflaton overdensities have already settled onto the so-called \textit{oscillon track}~\cite{Zhou:2024mea}, a quasi-attractor where all initial overdensities, even those initially deviating from the oscillon configuration, eventually converge.
To further confirm this behavior, we follow the evolution of the most energetic oscillon's profile from extraction until its decay (see Fig.~\ref{fig:energy-lifetime}). Then, at all times corresponding to a minimum of the kinetic energy, we fit the radial profile obtained from the 1+1 radial simulations to the Gaussian shape~\eqref{eq:Gaussian-fit}, finding, as expected, that the oscillon evolves approximately along the curve~\eqref{eq:ARfit}. It is worth noticing that slight mismatches at small amplitudes between Fig.~\ref{fig:AmpRad} and Fig.~\ref{fig:energy-lifetime} arise from the inherently different type of simulations being compared. Indeed, when fitting the profile from 3+1 lattice simulations, sphericity is an imposed condition and, while not widely spread, the field values at each radial distance do not perfectly coincide. Conversely, in the 1+1 radial simulation, sphericity is strictly imposed, leading to slight differences for the smaller oscillons. Nonetheless, as shown in Fig.~\ref{fig:energy-lifetime}, the evolution of the system is dominated by the few most energetic oscillons, which contain the majority of the energy.

%%%%%%%%%%%%%%%%%%%%%%%%%%%%%%%%%%%%%%%%%%%%%%%%%%%%%%%%%%%%%%%%%%%%%%%%
%%%%%%%%%%%%%%%%%%%%%%%%%%%%%%%%%%%%%%%%%%%%%%%%%%%%%%%%%%%%%%%%%%%%%%%%

\begin{figure*}
    \begin{center}
        \includegraphics[width=0.65\textwidth]{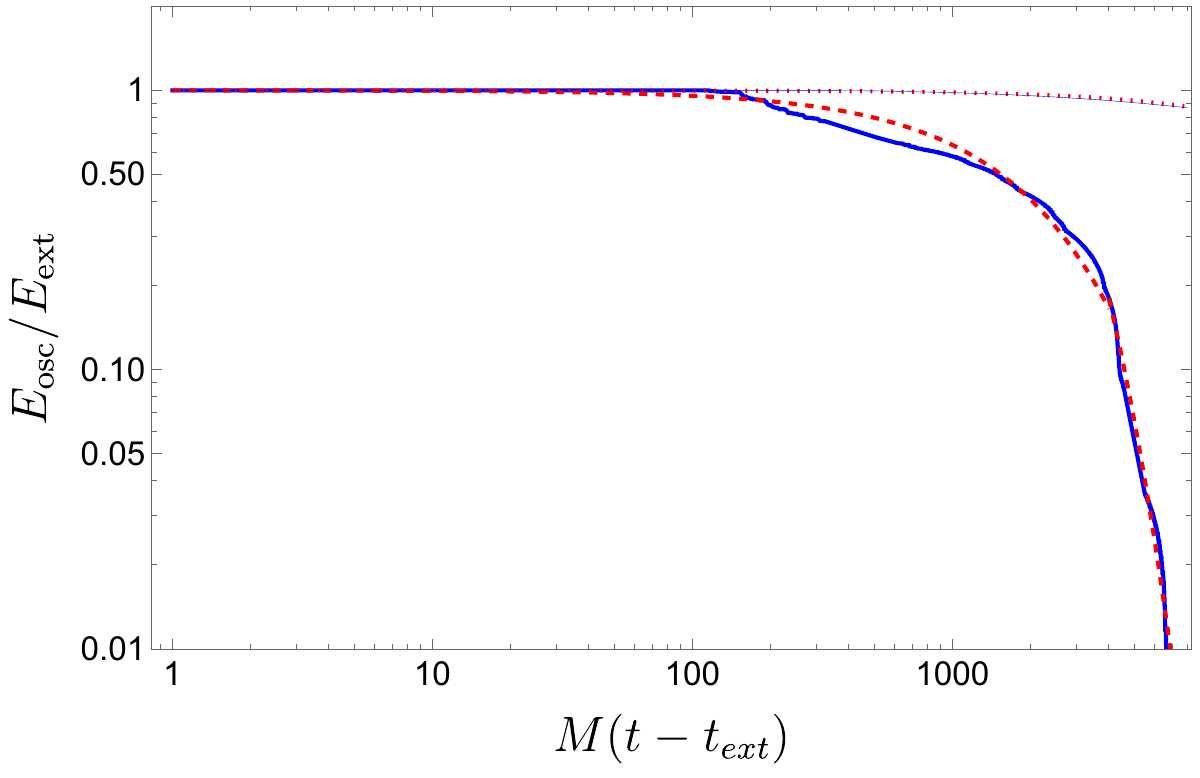} 
    \end{center}
    \caption{Evolution of the total energy stored in oscillons obtained from the radial simulations, for both the full potential with a quartic minimum \eqref{eq:canonical-field-potential}  (thick blue), and its corresponding quadratic approximation in the limit $\phi_c\rightarrow 0$ (thin blue). For each case we show a fit to the exponential template \eqref{eq:exp-approx} (red lines).}
    \label{fig:energy-loss}
\end{figure*}

We now turn to estimating the impact of the oscillons' evolution and decay on the post-inflationary evolution of the equation of state. Additionally, to better understand the role of the quartic potential at small field amplitudes, we compare our results with the case in which the potential has a purely quadratic minimum (corresponding to the formal limit $\phi_c \rightarrow 0$), while still assuming the same initial radial profile distribution. As illustrated in Fig.~\ref{fig:energy-loss}, the amount of radiated energy extracted from our simulations is noticeably smaller for the case with a purely quadratic potential. 

To determine the onset of radiation domination using our results, we approximate the energy evolution of oscillons as
\begin{equation}
    \label{eq:exp-approx}
    E_{\rm osc}(t>t_{\rm ext})=E_{\rm ext}  \,e^{-\Gamma\,( t-t_{\rm ext})}\,
    \end{equation}
with $E_{\rm ext}$ the total energy of oscillons at the extraction time and $\Gamma$ a constant decay rate. We then fit the exponential decay model in~\eqref{eq:exp-approx} to the simulation data, obtaining
\begin{equation}
    \label{eq:gamma-comparison}
    \Gamma_{\rm quartic}\simeq\begin{cases} 
5\cdot 10^{-4} M & \hspace{5mm} {\rm for} \hspace{5mm}t-t_{\rm ext}<4000 \\
10^{-3} M &\hspace{5mm}{\rm for} \hspace{5mm}t-t_{\rm ext}>4000
   \end{cases} \;, \hspace{10mm} \Gamma_{\rm quadratic}  \simeq 10^{-5}M\,.
\end{equation}

%%%%%%%%%%%%%%%%%%%%%%%%%%%%%%%%%%%%%%%%%%%%%%%%%%%%%%%%%%%%%%%%%%%%%%%%
%%%%%%%%%%%%%%%%%%%%%%%%%%%%%%%%%%%%%%%%%%%%%%%%%%%%%%%%%%%%%%%%%%%%%%%%

Using this information, we can estimate the time at which the Universe becomes radiation-dominated. Although our radial simulations do not incorporate the expansion of the Universe, oscillons behave as non-relativistic matter, being effectively decoupled from the cosmic expansion~\cite{Allahverdi:2020bys} and maintaining a fixed physical size. As a result, their energy density evolves as 
\begin{equation}
    \label{eq:osc-density}
    \rho_{\rm osc}(t>t_{\rm ext})=\frac{E_{\rm osc}}{V}=\frac{E_{\rm ext}  \,e^{-\Gamma\,( t-t_{\rm ext})}}{a^3(t) \mathcal{V}_{\rm ext}}\,
\end{equation}
with $\mathcal{V}_{\rm ext}$ the comoving volume of the 3-dimensional lattice at the extraction time $t_{\rm text}$. By differentiating \eqref{eq:osc-density} with respect to time, we identify two distinct contributions to the decrease in the oscillons' energy density: the expansion of the Universe and the energy radiated by them. Assuming that the radiated energy behaves as a relativistic fluid, the evolution of the non-relativistic matter (oscillons) and radiation densities can be determined by solving the following system of coupled differential equations, 
\begin{equation}
 \label{eq:energy-tranfer}
        \dot{\rho}_{\rm osc}=-3 H \rho_{\rm osc}- \Gamma \rho_{\rm osc}\,, \hspace{10mm}
        \dot{\rho}_{\rm rad}=-4 H \rho_{\rm rad}+ \Gamma \rho_{\rm osc}\,. 
\end{equation}
From these expressions, we deduce that efficient energy transfer between matter and radiation requires the condition $\Gamma \sim H$ to be satisfied. Indeed, as long as the Hubble rate $H$ is much larger than the transfer rate $\Gamma$, the expansion dilutes any radiation-like component faster than matter. However, as $H$ decreases with the expansion, eventually we reach a point where $\dot{\rho}_{\rm osc}\sim -4H \rho_{\rm osc}$, making the energy transfer between the two fluids efficient. In our case, this process is already efficient at the extraction time, as $\Gamma>H_{\rm ext}$. 

For the initial energy densities, we adopt values obtained from the 3+1 lattice simulations at the extraction time, when approximately 75\% of the total energy is stored in oscillons, such that $\rho_{\rm osc} (t_{\rm ext})=(3/4)\rho_{\rm tot}^{\rm ext}$ and $\rho_{\rm rad} (t_{\rm ext}) =(1/4)\rho_{\rm tot}^{\rm ext}$. With these initial conditions, we numerically solve the system of equations~\eqref{eq:energy-tranfer} to determine the evolution of the oscillons' and radiation densities. Using the resulting  time dependence of energy densities, we calculate the evolution of the equation-of-state parameter and the scale factor
\begin{equation}
    \label{eq:eos-evolution}
    w=-1-\frac{\dot{\rho}}{3 H \rho}\,,\hspace{10mm} \frac{\dot{a}}{a}=\sqrt{\frac{\rho}{3}}\,,
\end{equation}
finding that the Universe transitions into a radiation-dominated stage \footnote{We define the onset of radiation domination as the moment when $\rho_{\rm rad}=0.99\,\rho_{\rm tot}$.} $\Delta \mathcal{N}_{\rm end}^{\rm rd}=3.81$ $e$-folds after the end of inflation.
These results contrast with the commonly reported behavior of oscillons in the literature, where these objects are typically long-lived, being able to survive for $\mathcal{O}(10^8)$ oscillations~\cite{Zhang:2020bec}. In those scenarios, the oscillon's stability is granted by the fact that at small field values, therefore far from the center, the potential is essentially quadratic, resembling a free theory. Consequently, radiation is predominantly emitted from the core, where the field amplitude is largest and explores the non-linear regime of the potential. In our case, however, the minimum of the potential is quartic, leading to a notably different picture. While, as in the standard case, some radiation still originates from the core of the oscillon, there is an additional contribution from the external layer. In particular, at large $r$, the field amplitude is always below the crossover scale $\phi < \phi_c$, causing the outer layers of the oscillon to continuously radiate and gradually decrease in amplitude. As illustrated in Fig.~\ref{fig:energy-loss}, this reduction in amplitude at the outer layers introduces a gradient that drives the oscillon to shrink, significantly shortening its lifetime as compared to the case of a purely quadratic minimum. 

%%%%%%%%%%%%%%%%%%%%%%%%%%%%%%%%%%%%%%%%%%%%%%%%%%%%%%%%%%%%%%%%%%%%%%%%
\begin{figure*}
    \begin{center}
       \includegraphics[width=0.65\textwidth]{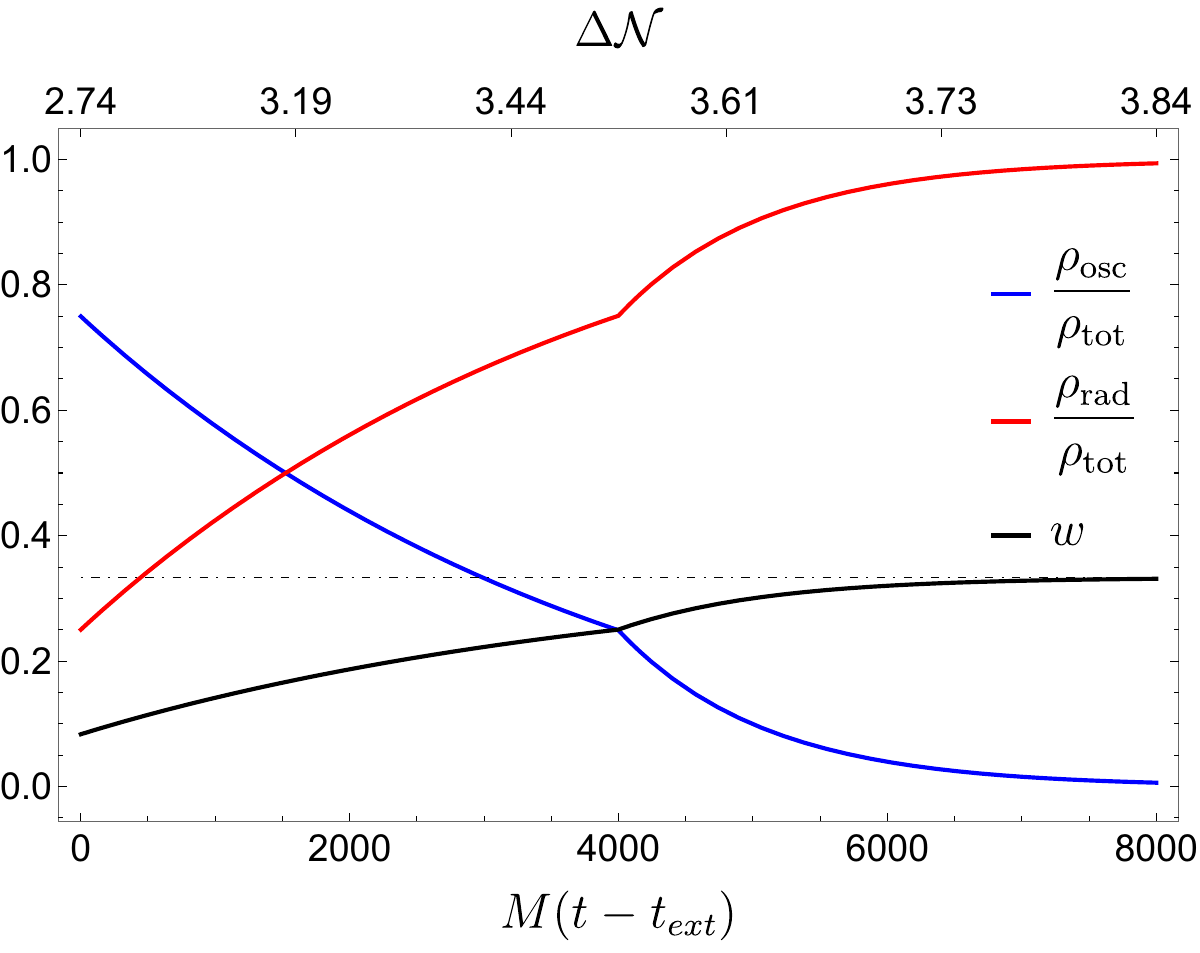} \,\,
    \end{center}
    \caption{Time evolution of the energy density for oscillons and radiation, as well as the equation of state parameter, using the system of equations~\eqref{eq:energy-tranfer}. The dashed-dotted line corresponds to the value $w=1/3$. As the decay rate is already larger than the initial Hubble rate, the transfer between oscillons and radiation is already efficient at $t_{\rm ext}$. The system enters the stage of radiation domination at $Mt\simeq 7000$. The upper x-axis shows the corresponding number of $e$-folds after the end of inflation.}
    
    \label{fig:radiation-domination}
\end{figure*}
%%%%%%%%%%%%%%%%%%%%%%%%%%%%%%%%%%%%%%%%%%%%%%%%%%%%%%%%%%%%%%%%%%%%%%%%

%%%%%%%%%%%%%%%%%%%%%%%%%%%%%%%%%%%%%%%%%%%%%%%%%%%%%%%%%%%%%%%%%%%%%%%%
\subsection{Inflationary observables and gravitational waves signatures}\label{sec:results}
%%%%%%%%%%%%%%%%%%%%%%%%%%%%%%%%%%%%%%%%%%%%%%%%%%%%%%%%%%%%%%%%%%%%%%%%

Equipped with the knowledge of the post-inflationary evolution, and in particular its duration and equation-of-state parameter, we can now investigate how this influence measurable quantities, such as the inflationary observables and the amplitude and peak frequency of the GWs spectrum produced by the oscillons. Let us focus first on the impact on inflationary observables. Specifically, we aim to determine the exact amount of $e$-folds of inflation required to solve the horizon problem, i.e. to match the pivot scale of the CMB with the current Hubble horizon, while traversing through the different stages of the cosmic history, namely
\begin{equation}
    \label{eq:match-scales}
    \frac{a_*H_*}{a_0H_0}=e^{-\mathcal{N_*}}\frac{a_{\rm end}}{a_{\rm rd}}\frac{a_{\rm rd}}{a_{\rm eq}}\frac{a_{\rm eq}}{a_0}\frac{H_*}{H_{\rm eq}}\frac{H_{\rm eq}}{H_0}\,,
\end{equation}
with the subscripts indicating the times at which every quantity is evaluated, namely: horizon crossing ($*$), end of inflation (end), onset of radiation domination (rd), matter-radiation equality (eq) and the present time (0). Using the current Planck 2018 constraints~\cite{Planck:2018jri} on the Hubble parameter $H_0$ and matter density parameter $\Omega_{\rm m,0}$, we can rewrite Eq.~\eqref{eq:match-scales} as\footnote{For the sake of clarity we have reintroduced the Planck mass.}
\begin{equation}
    \label{eq:correct-efolds-generic}
    \mathcal{N}_*=67-\ln\left(\frac{k_*}{a_0H_0}\right)+\frac{1}{4} \ln \left( \frac{V_*^2}{M_P^4 \;\rho_{\rm rd}}\right)-\Delta\mathcal{N}_{\rm end}^{\rm rd}\,.
\end{equation}
To separate the contributions coming from inflationary and post-inflationary evolution, we introduce the time-averaged equation-of-state parameter between times $t_1$ and $t_2$, 
\begin{equation}
    \label{eq:eos-average} \bar{w}(t_1,t_2)=\frac{1}{t_2-t_1}\int_{t_1}^{t_2}w(t)dt\,.
\end{equation}
Therefore, by fixing the pivot scale to $k_*=0.05  \;\rm Mpc^{-1}$, we eventually arrive to the following expression for the number of $e$-folds
\begin{equation}
    \label{eq:e-folds}
    \mathcal{N}_*=61.5+\frac{1}{4} \ln \frac{V^2_*}{M_P^4 \rho_{\rm end}}+\frac{3 \bar{w}(t_{\rm end},t_{\rm RD})-1}{4}\Delta\mathcal{N}^{\rm RD}_{\rm end}\,, 
\end{equation}
with the three terms representing the respective contributions from the standard cosmic history, the inflationary parameters, and the post-inflationary evolution. At this point, it is important to highlight that the standard approach to determining the parameters in the inflationary potential~\eqref{eq:canonical-field-potential} involves selecting a reference value for $\mathcal{N}_*$ and adjusting the parameters accordingly through the relation $ \mathcal{N}=\int_{\phi_{\rm end}}^\phi (2 \epsilon(\phi'))^{-1/2}d \phi'$. Once this is done, two procedures can be employed:
\begin{enumerate}
    \item Determine, either analytically or numerically, which set of parameters results in the same $\mathcal{N}_*$ when using both Eq.~\eqref{eq:e-folds}
 and Eq.~\eqref{eq:e-folds0};
 \item Start with a fiducial $\mathcal{N}^{(0)}_*$ to fix the parameters of the potential, determine the post-inflationary evolution, and use Eq.~\eqref{eq:e-folds} to compute a new $\mathcal{N}_*^{(1)}$. This process can then be iterated until the difference $|\mathcal{N}_*^{(p)}-\mathcal{N}_*^{(p+1)}|$ is smaller then the desired precision. 
 \end{enumerate}
 The procedure (1), although more rigorous, requires some sort of parametric description of $\bar{w}(t_{\rm end},t_{\rm RD})$ and $\Delta\mathcal{N}^{\rm RD}_{\rm end}$, which may be difficult to obtain for highly non-linear models that require 3+1 lattice simulations. On the other hand, the procedure (2) assumes that the iterative process converges, but is achievable with numerical simulations, and has been shown to converge within few iterations in some specific scenarios~\cite{Antusch:2021aiw}. In our case, combining Eq.~\eqref{eq:e-folds0} with the values of the parameters of our benchmark scenario $\{\lambda, \xi, c\} = \{0.001, 5 \cdot 10^4, 1.21 \cdot 10^7\}$, we obtain $\mathcal{N}_*^{(0)}=55$. Given that the iterating procedure (2) introduces some arbitrariness in terms of which of the three model parameters to modify, and since their change could affect the post-inflationary physics through shifts in the ratios between the scales $(\phi_{\rm end}, \phi_i,\phi_c)$, we restrict ourselves to the first step of the iterative procedure. After combining the 3+1 lattice simulations with the 1+1 ones, we refine our prediction for the spectral tilt,
\begin{equation}
    \label{eq:observables-corrected}
    \mathcal{N}^{(1)}_*\simeq53\hspace{5mm}\longrightarrow \hspace{5mm} n_s=1-\frac{2}{\mathcal{N}_*^{(1)}}\simeq0.9623\,,
\end{equation}
which, assuming the current central value, falls within the $1\sigma$ interval of the forecasted sensitivity for CMB Stage-IV experiments~\cite{LiteBIRD:2022cnt}, $n_s\in [0.9544,0.9674]$. 

Another important observable byproduct of a modified post-inflationary expansion in models supporting the formation of oscillons is the presence of a stochastic background of gravitational waves. In fact, anisotropies in the inflaton configuration can source transverse-traceless (TT) linear metric perturbations $h_{ij}$ on top of the FLRW metric
\begin{equation}
    \label{eq:metric-perturbation}
    ds^2=-dt^2 +a^2(t)\left[\delta_{ij}+h_{ij} \right]dx^idx^j\,, \hspace{10mm}\partial_ih_{ij}=0\,,\hspace{5mm} h_{ii}=0\,,
\end{equation}
whose dynamics is described by the wave equation
\begin{equation}
    \label{eq:GW-KG}
    \ddot{h}_{ij}+3 H \dot{h}_{ij}-a^{-2}\nabla^2 h_{ij}=2 a^{-2}\Pi^{\rm TT}_{ij}\,, 
\end{equation}
with $\Pi^{\rm TT}_{ij}=(\partial_i \phi \partial_j \phi)^{\rm TT}  $ the transverse-traceless~(TT) part of the associated anisotropic energy-momentum tensor. As already done in~\cite{Piani:2023aof}, we make use of the GWs module implemented in \CL\cite{tech} to extract the energy density power spectrum of the GWs signal \cite{Figueroa:2020rrl}, 
\begin{equation}
    \label{eq:GWPS}
    \Omega_{\rm GW}=\frac{1}{3 H^2} \frac{{\rm d}\rho_{\rm GW}}{{\rm d}\log \,k}~, \hspace{10mm}{\rm with}\hspace{10mm} \rho_{\rm GW}=\frac{1}{4}\langle \dot{h}_{ij} \dot{h}_{ij}\rangle~.
\end{equation}
As noted in~\cite{Piani:2023aof}, the GWs spectrum grows initially with time, developing a peak within the momentum band undergoing tachyonic growth. When fragmentation begins, the spectrum grows rapidly until oscillons form and reach a quasi-spherical shape, when the growth halts due to the absence of any significant quadrupole moment. In order to connect the final GWs spectrum at the time of emission ($t_e$) to that at present times, we need to account for the post-inflationary evolution up to the onset of radiation domination. By assuming an adiabatic evolution of the Universe up to the present day, i.e. ensuring entropy conservation, we can appropriately rescale the amplitude $\Omega_{0,{\rm GW}}$ and frequency $f_0$ of the spectrum, obtaining~\cite{Dufaux:2007pt}
\begin{eqnarray} \label{eq:peak-frequency}
&& f_0=\frac{k}{2\pi a_0}= 4 \times 10^{10} \dfrac{ k}{a_{\rm e} \rho_{\rm e}^{1/4}}\exp\left[{{-\frac{1}{4}(1-3\bar w(t_{\rm e},t_{\rm RD}))\Delta {\cal N}_{\rm e}^{\rm RD} }}\right]\,~\rm Hz\,, \\
&& \Omega_{0,{\rm GW}}~h^2 \simeq 1.6 \times 10^{-5} \exp\left[{-(1-3\bar w(t_{\rm e},t_{\rm RD})))\Delta {\cal N}^{RD}_{\rm e}}\right] \, \Omega_{\rm e,GW}\,,
\end{eqnarray}
with $a_0$ the scale factor today, $h\sim0.7$  the uncertainty on the current value of the Hubble parameter and the quantities $a_{\rm e}$, $\rho_{\rm e}$ and $\Omega_{\rm e,GW}$ denoting respectively the scale factor, the total energy density and the gravitational waves' energy fraction at the time of emission.
\begin{figure*}
    \begin{center}
       \includegraphics[width=0.65\textwidth]{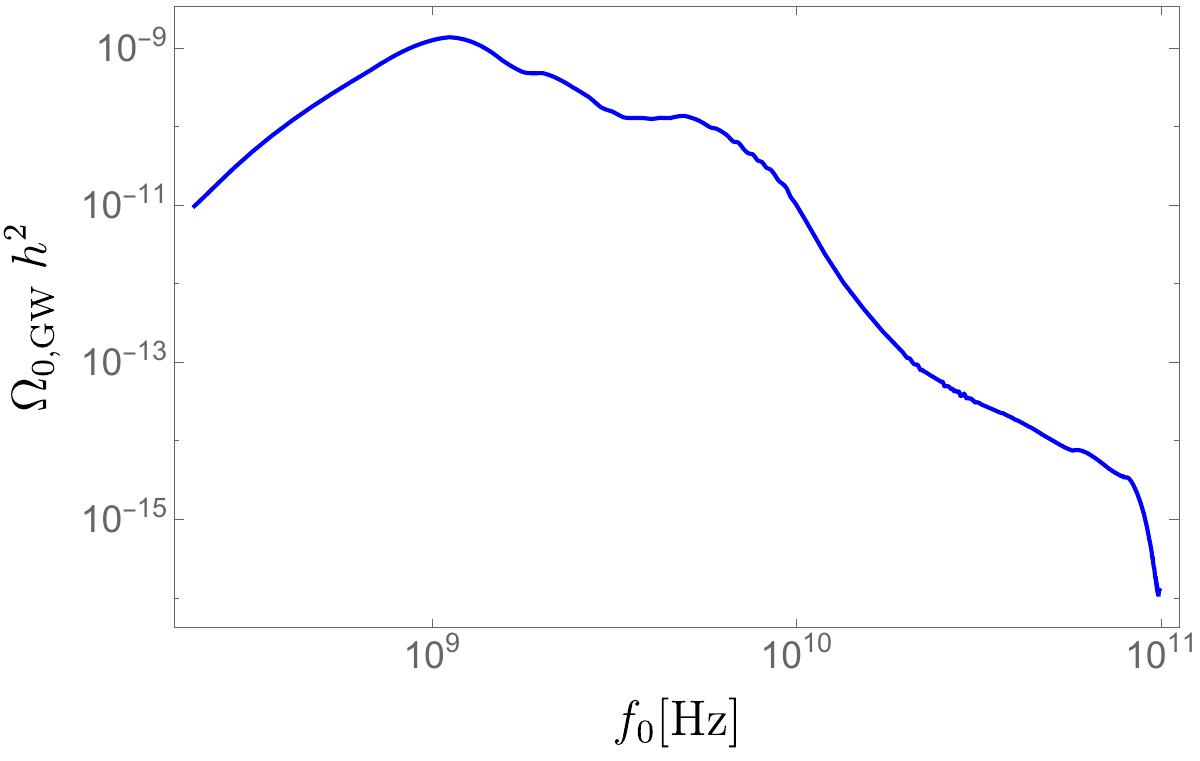} \,\,
    \end{center}
    \caption{Spectrum of the SGWB today. Despite the large amplitude of the signal, the peak of the spectrum is approximately at 1 GHz, far away from the experimental windows available with current detectors. The spectrum shown is obtained from a simulation with $N^3=512^3$ lattice points, and a box with side length $L=30 M^{-1}$.}
    \label{fig:GW}
\end{figure*}
The resulting spectrum, shown in Fig.~\ref{fig:GW}, exhibits an amplitude commensurable with the sensitivity of current GWs experiments. However, similarly to other preheating scenarios, the peak frequency lies around $f_{\rm peak}\sim\mathcal{O}({\rm GHz})$. While the frequency range of current experiments is several orders of magnitude lower, numerous proposals are being developed to probe ultra-high frequency signals from the early Universe in the  MHz-GHz range~\cite{Aggarwal:2020olq, Herman:2022fau, Aggarwal:2025noe}.

%%%%%%%%%%%%%%%%%%%%%%%%%%%%%%%%%%%%%%%%%%%%%%%%%%%%%%%%%%%%%%%%%%%%%%%%
\section{Summary and discussion} \label{sec:conclusions}
%%%%%%%%%%%%%%%%%%%%%%%%%%%%%%%%%%%%%%%%%%%%%%%%%%%%%%%%%%%%%%%%%%%%%%%%

In this paper, we have presented a detailed study of the preheating stage in a non-minimally coupled inflaton model formulated within the framework of Einstein-Cartan gravity. Our primary focus was to investigate the formation, evolution, and eventual decay of oscillons, with the aim of assessing their influence on the onset of radiation domination.

Focusing on models that involve polynomial operators of dimension $d \leq 4$ and a scale-invariant potential, we identified a specific class of scenarios leading to an effective Einstein-frame potential exhibiting three distinct behaviors: a plateau at large field values, a quadratic region at intermediate scales, and a quartic regime at small field values. The interplay between the first two regions governs the initial dynamics of the preheating stage, ensuring that during the first few oscillations, the inflaton field traverses the tachyonic region of the potential, leading to an exponential growth of infrared modes.

To investigate the formation of oscillons in this setting, we made use of 3+1 classical lattice simulations, a numerical method inherently constrained to the early stages of oscillon evolution, due to two significant factors that compromise its reliability over time: i) the fixed physical size of oscillons, which leads to a loss of resolution as the Universe expands and ii) the influence of the quartic region of the potential, which progressively amplifies UV modes through re-scattering processes. To address these limitations and extend our analysis to later times, we concentrated on the evolution of individual oscillons after they had stabilized into quasi-spherical configurations. By extracting the field profiles from the preceding 3+1 lattice simulations, we identified their essential structural features, modeling their radial profiles as Gaussian distributions. These distributions were then employed as initial conditions for subsequent 1+1 radial simulations, from which we determined the oscillon decay rate, the effective equation of state governing the system from formation to complete decay, the precise number of $e$-folds required to address the flatness and horizon problems in this scenario, the key inflationary observables, and the amplitude and frequency of the gravitational wave spectrum generated at oscillon formation. Overall, we found that, assuming the current central value, the spectral tilt of primordial density perturbation falls within the $1\sigma$ interval projected for the upcoming LiteBIRD experiment \cite{LiteBIRD:2022cnt}. Meanwhile, the peak of the gravitational wave spectrum is located at $f \sim \mathcal{O}(\rm GHz)$, i.e. beyond the reach of current detectors but well within the sensitivity range of proposed ultra-high frequency detectors \cite{Aggarwal:2020olq,Herman:2022fau,Aggarwal:2025noe}.

While this work represents an initial step toward understanding the preheating dynamics in these settings, there are several aspects that our analysis did not fully capture, as well as approximations employed, which are important to note:
\begin{enumerate}
    \item \textit{Quantum effects:} In this study, we have focused exclusively on a classical treatment of the preheating process, neglecting potential quantum effects. However, existing literature suggests that quantum effects could play a significant role on oscillons decay, potentially leading to a substantial reduction in the oscillon lifetime~\cite{Hertzberg:2010yz}. Incorporating quantum corrections into the analysis would likely provide a more complete understanding of the oscillon dynamics, particularly at shorter timescales.   
    \item \textit{Deviation from spherical symmetry:} Our treatment is based on the assumption that oscillons can eventually be approximated as spherically symmetric objects, with their profiles extractable at arbitrary distances. This assumption is qualitatively supported by  our 3+1 dimensional simulations, showing that as time progresses, the overdensities tend to evolve into an approximately spherical shape. However,  radial symmetry can only be assumed within the oscillon itself, and our 1+1 dimensional simulations extend over distances much larger than those of the overdensities. Nevertheless, we have verified that small modifications to the radial tails of the initial profiles do not significantly impact the oscillons' lifetime, particularly for the more energetic oscillons governing the system's evolution. Moreover, since oscillons are attractor solutions, even if the initial profile slightly deviates from the attractor, the system quickly adjusts it and follows the expected oscillon trajectory.
    \item \textit{Interactions with the environment:} In our approach, the late-time behavior of the system is modeled by treating the oscillons as isolated objects, thus neglecting any potential interactions between them that could lead to deformations or subtle changes in the energy evolution. Furthermore, as discussed in Section \ref{sec:OscLate}, oscillons radiate from their external shells whenever the radial field value falls below the crossover scale. This suggests that the residual inflaton condensate not incorporated into oscillons could a priori influence large distances, potentially altering the dynamics of the external shells. However,  it is important to keep in mind that by the time of extraction, $75\%$ of the total energy is contained within the oscillons, with the most energetic ones having an energy density significantly higher than average. As a result, we anticipate that this effect will not substantially affect our findings.
    \item \textit{Gravitational effects:} Oscillon collective gravitational interactions could potentially lead to the formation of primordial black holes and clustering under suitable conditions, \cite{Cotner:2018vug,Amin:2019ums}, a possibility not accounted for in our simulations, where metric perturbations are completely disregarded. The typical oscillon sizes observed in our simulations make, however, unlikely that any of them would satisfy the conditions for collapse.
    \item \textit{Second-order gravitational waves: } In this work, we have focused solely on the gravitational waves generated during the inflaton fragmentation and oscillon formation phases. However, it is also plausible that oscillons may produce additional gravitational wave backgrounds during their decay \cite{Lozanov:2023aez, Lozanov:2023knf}. The resulting signal would mainly depend on the asymmetry that the oscillons acquire during this process, as well as the distribution of their lifetimes.
\end{enumerate} 
In spite of the specific focus of this study, the general principles and methods we used can be extended to a variety of inflationary models and cosmological settings such as non-canonical inflationary scenarios and $\alpha$-attractor settings \cite{Kallosh:2013hoa,Kallosh:2013yoa,Galante:2014ifa,Carrasco:2015pla,Carrasco:2015rva,Kallosh:2015lwa,Artymowski:2016pjz}. Furthermore, although our study has focused on a scalar-tensor implementation of Einstein-Cartan gravity without incorporating other matter degrees of freedom, the results can be partially applied to models featuring additional interactions. A notable case in point is the Einstein-Cartan embedding of the well-known Higgs inflation scenario, where the Standard Model Higgs field is directly identified with the inflaton condensate. Indeed, as discussed in Ref.~\cite{Piani:2023aof}, the formation of oscillons in this framework
is largely unaffected by the couplings of the Higgs-inflaton field to the Standard Model species. Before fragmentation takes place the gauge bosons can be approximate as three scalar degrees of freedom with a mass that varies during the oscillations of the Higgs condensate~\cite{Piani:2023aof,Figueroa:2015rqa,Enqvist:2015sua}. As the background evolves towards the minimum of the potential their mass evolves non-adiabatically, causing an increase in their number density, which is further boosted at every semi-oscillation by a Bose-enhancement mechanism. On the other hand, at large field values their mass increases, making their perturbative decay into SM fermions efficient enough, so that their density is partially depleted, slowing down the Bose-enhancement process.  As a result, the energy transfer to fermions via gauge boson decay typically requires hundreds of oscillations to become significant~\cite{Garcia-Bellido:2008ycs,Bezrukov:2008ut,Repond:2016sol} while the oscillons are typically forming within $\mathcal{O}(10)$ oscillations. Once the oscillons have formed the Higgs has an inhomogeneous distribution, which does not allow us to employ the previous formalism. In this context we might naively expect that the vectorial nature of gauge bosons will cause their production from a spherically symmetric structure to be suppressed by multipoles, when compared to the radiation of Higgs quanta.\footnote{However, the situation might change for generic models in which the inflaton is coupled to one or more daughter fields, potentially affecting their lifetime~\cite{Shafi:2024jig}.} Consequently, our results provide an upper bound on the lifetime and energy release of oscillons, establishing a benchmark for future investigations properly incorporating the full spectrum of Standard Model interactions. 

%%%%%%%%%%%%%%%%%%%%%%%%%%%%%%%%%%%%%%%%%%%%%%%%%%%%%%%%%%%%%%%%%%%%%%%%
\acknowledgments
%%%%%%%%%%%%%%%%%%%%%%%%%%%%%%%%%%%%%%%%%%%%%%%%%%%%%%%%%%%%%%%%%%%%%%%%

 JR (ORCID 0000-0001-7545-1533) is supported by a Ramón y Cajal contract of the Spanish Ministry of Science and Innovation with Ref.~RYC2020-028870-I. This work was supported by the project PID2022-139841NB-I00 of MICIU/AEI/10.13039/501100011033 and FEDER, UE.
 FT (ORCID 0000-0003-1883-8365) is supported by a \textit{Beatriu de Pinós} fellowship (reference number: 2022 BP 00063) from the Ministry of Research and Universities of the Government of Catalonia, and partially supported by grants PID2019-108122GB-C32 from the Spanish Ministry of Science and Innovation, Unit of Excellence Maria de Maeztu 2020-2023 of ICCUB (CEX2019-000918-M) and AGAUR 2021 SGR 00872.
 MP (ORCID ID 0000-0002-2387-5948) acknowledges the Funda\c c\~ao para a Ci\^encia e a Tecnologia (FCT), Portugal, for the financial support to the Center for Astrophysics and Gravitation-CENTRA, Instituto Superior T\'ecnico,  Universidade de Lisboa, through the Project No.~UIDB/00099/2020.  MP thanks  also the support of this agency through the Grant No. SFRH/BD/151003/2021 in the framework of the Doctoral Program IDPASC-Portugal. MP also wishes to thank Giorgio Laverda and Joan Bachs Esteban for useful discussions and suggestions during the preparation of this work. The authors are also grateful to Ioannis Gialamas for useful comments on vanishing torsion terms.

\appendix

%%%%%%%%%%%%%%%%%%%%%%%%%%%%%%%%%%%%%%%%%%%%%%%%%%%%%%%%%%%%%%%%%%%%%%%%
\section{Details on radial simulations} \label{sec:rad-sim}
%%%%%%%%%%%%%%%%%%%%%%%%%%%%%%%%%%%%%%%%%%%%%%%%%%%%%%%%%%%%%%%%%%%%%%%%

\begin{figure*}
\begin{center}
    \includegraphics[width=0.65\textwidth]{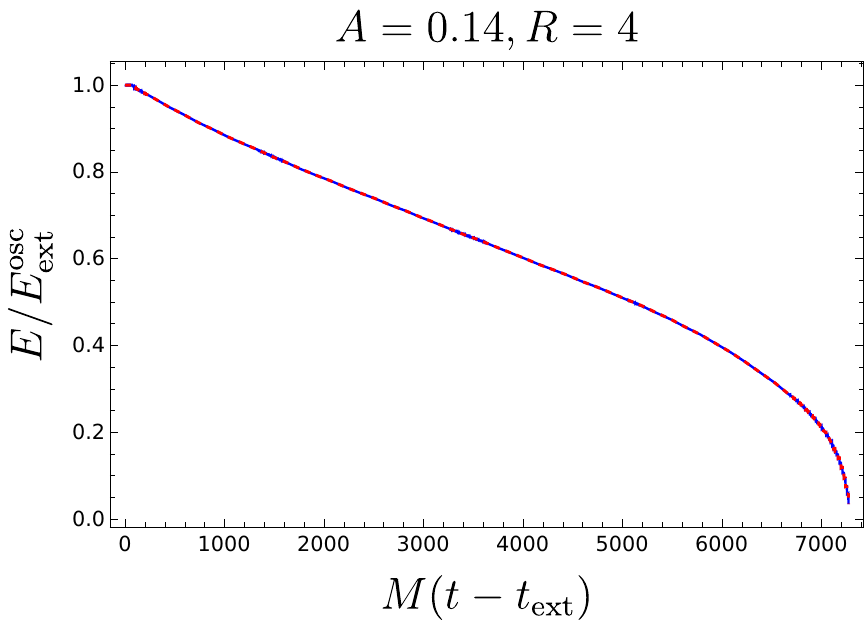}
\end{center}
\caption{Tracking of the energy conservation for a simulation with $A=0.14$, $R=4$. The blue line shows the total energy in the lattice as a function of time, which decreases as we continuously truncate the scalar waves at large radial distances. The red line is the initial energy in the box minus the accumulated subtracted energy due to the truncation technique.}
\label{fig:Econ}
\end{figure*}

In this Appendix, we provide details on the numerical algorithm used to solve the oscillons dynamics in 1+1 dimensions in Section \ref{sec:OscLate}. Our approach consist in solving Eq.~\eqref{eq:rad-eom1} in a one-dimensional lattice with periodic boundary conditions of length $L$ and number of points $N$. The nodes of the lattice are defined at the spatial coordinates $r = (n + 1/2) \Delta x$, with $n =1, 2, \dots N$, where $\Delta x \equiv N /L$ is the lattice spacing. The results of this paper are obtained with the choice $L = 100 R_{i}$ and $N= 50000$, which gives a lattice spacing of $\Delta x= 500^{-1}R_i$. More specifically, we solve Eq.~\eqref{eq:rad-eom1} with a staggered leapfrog algorithm with the time step fixed to $\Delta t= 0.5 \Delta x$. The spatial continuous derivatives are approximated by the following fourth-order accurate expressions \cite{DiscreteDerivatives},
\begin{align}
\frac{\partial \phi}{\partial r} \,&\simeq\, \frac{\phi(r-2\Delta {\rm x} ) - 8\phi (r-\Delta {\rm x} ) +8\phi(r+\Delta {\rm x} )  -\phi (r+2\Delta {\rm x} )}{12\,\Delta {\rm x}}  \,, \\
\frac{\partial^2\phi}{\partial r^2}\,&\simeq \,\frac{-\phi (r+2\Delta {\rm x} ) + 16\phi (r+\Delta {\rm x} )  -30\phi (r ) +16\phi (r-\Delta {\rm x} ) -\phi (r-2\Delta {\rm x} )}{12\,\Delta {\rm x}^2} \,.  \,\,\,\,\,\,\,\,
\end{align}
As the oscillon oscillates, it emits field waves that eventually arrive at the rightmost boundary of the lattice. In order to avoid this problem, the field amplitude is reset each time a wave arrives to the boundary as 
\begin{equation}
\phi(t,r) \,=\, \begin{cases}
\phi(t,r)&\quad{\rm for}\quad r< r_{_{\hspace{-0.06cm} \Delta}}- \Delta_R \,,\\
\phi(t,r) \mathcal{G}(r;r_{_{\hspace{-0.06cm} \Delta}})&\quad{\rm for}\quad r_{_{\hspace{-0.06cm} \Delta}}-\Delta_R \le r\le r_{_{\hspace{-0.06cm} \Delta}}+\Delta_R \,,\\
0&\quad{\rm for}\quad r>r_{_{\hspace{-0.06cm} \Delta}}+\Delta_R \,,
\end{cases}
\end{equation}
with  
\begin{equation} \mathcal{G}(r;r_{_{\hspace{-0.06cm} \Delta}})\,\equiv\,\frac{e^{-\gamma(r-r_{_{\hspace{-0.06cm} \Delta}})}}{1+e^{-\gamma(r-r_{_{\hspace{-0.06cm} \Delta}})}}\ , \end{equation}
a truncation function. A similar truncation procedure is applied to the scalar field's time-derivative $\phi'$. For the truncation function, we choose the specific parameters 
$\Delta_R = 10 R$, $\gamma = 10 $ and $r_{_{\hspace{-0.06cm} \Delta}}=60R$, 
but we have checked that varying these within reasonable ranges does not modify the measured lifetimes for the oscillons. Finally, in order to assess the accuracy of our simulation, we track the energy conservation within our lattice, see Fig.~\ref{fig:Econ}.
%%%%%%%%%%%%%%%%%%%%%%%%%%%%%%%%%%%%%%%%%%%%%%%%%%%%%%%%%%%%%%%%%%%%%%%%%%%%%%%%
\bibliographystyle{JHEP}
\bibliography{mybib,extra}

\providecommand{\href}[2]{#2}\begingroup\raggedright\begin{thebibliography}{100}

\bibitem{Planck:2018jri}
{\scshape Planck} collaboration, \emph{{Planck 2018 results. X. Constraints on inflation}}, \href{https://doi.org/10.1051/0004-6361/201833887}{\emph{Astron. Astrophys.} {\bfseries 641} (2020) A10} [\href{https://arxiv.org/abs/1807.06211}{{\ttfamily 1807.06211}}].

\bibitem{BICEP:2021xfz}
{\scshape BICEP, Keck} collaboration, \emph{{Improved Constraints on Primordial Gravitational Waves using Planck, WMAP, and BICEP/Keck Observations through the 2018 Observing Season}}, \href{https://doi.org/10.1103/PhysRevLett.127.151301}{\emph{Phys. Rev. Lett.} {\bfseries 127} (2021) 151301} [\href{https://arxiv.org/abs/2110.00483}{{\ttfamily 2110.00483}}].

\bibitem{Guth:1980zm}
A.H.~Guth, \emph{{The Inflationary Universe: A Possible Solution to the Horizon and Flatness Problems}}, \href{https://doi.org/10.1103/PhysRevD.23.347}{\emph{Phys. Rev. D} {\bfseries 23} (1981) 347}.

\bibitem{Linde:1981mu}
A.D.~Linde, \emph{{A New Inflationary Universe Scenario: A Possible Solution of the Horizon, Flatness, Homogeneity, Isotropy and Primordial Monopole Problems}}, \href{https://doi.org/10.1016/0370-2693(82)91219-9}{\emph{Phys. Lett. B} {\bfseries 108} (1982) 389}.

\bibitem{Mukhanov:1981xt}
V.F.~Mukhanov and G.V.~Chibisov, \emph{{Quantum Fluctuations and a Nonsingular Universe}}, {\emph{JETP Lett.} {\bfseries 33} (1981) 532}.

\bibitem{Bassett:2005xm}
B.A.~Bassett, S.~Tsujikawa and D.~Wands, \emph{{Inflation dynamics and reheating}}, \href{https://doi.org/10.1103/RevModPhys.78.537}{\emph{Rev. Mod. Phys.} {\bfseries 78} (2006) 537} [\href{https://arxiv.org/abs/astro-ph/0507632}{{\ttfamily astro-ph/0507632}}].

\bibitem{Allahverdi:2010xz}
R.~Allahverdi, R.~Brandenberger, F.-Y.~Cyr-Racine and A.~Mazumdar, \emph{{Reheating in Inflationary Cosmology: Theory and Applications}}, \href{https://doi.org/10.1146/annurev.nucl.012809.104511}{\emph{Ann. Rev. Nucl. Part. Sci.} {\bfseries 60} (2010) 27} [\href{https://arxiv.org/abs/1001.2600}{{\ttfamily 1001.2600}}].

\bibitem{Amin:2014eta}
M.A.~Amin, M.P.~Hertzberg, D.I.~Kaiser and J.~Karouby, \emph{{Nonperturbative Dynamics Of Reheating After Inflation: A Review}}, \href{https://doi.org/10.1142/S0218271815300037}{\emph{Int. J. Mod. Phys. D} {\bfseries 24} (2014) 1530003} [\href{https://arxiv.org/abs/1410.3808}{{\ttfamily 1410.3808}}].

\bibitem{Kawasaki:1999na}
M.~Kawasaki, K.~Kohri and N.~Sugiyama, \emph{{Cosmological constraints on late time entropy production}}, \href{https://doi.org/10.1103/PhysRevLett.82.4168}{\emph{Phys. Rev. Lett.} {\bfseries 82} (1999) 4168} [\href{https://arxiv.org/abs/astro-ph/9811437}{{\ttfamily astro-ph/9811437}}].

\bibitem{Kawasaki:2000en}
M.~Kawasaki, K.~Kohri and N.~Sugiyama, \emph{{MeV scale reheating temperature and thermalization of neutrino background}}, \href{https://doi.org/10.1103/PhysRevD.62.023506}{\emph{Phys. Rev. D} {\bfseries 62} (2000) 023506} [\href{https://arxiv.org/abs/astro-ph/0002127}{{\ttfamily astro-ph/0002127}}].

\bibitem{Hannestad:2004px}
S.~Hannestad, \emph{{What is the lowest possible reheating temperature?}}, \href{https://doi.org/10.1103/PhysRevD.70.043506}{\emph{Phys. Rev. D} {\bfseries 70} (2004) 043506} [\href{https://arxiv.org/abs/astro-ph/0403291}{{\ttfamily astro-ph/0403291}}].

\bibitem{Hasegawa:2019jsa}
T.~Hasegawa, N.~Hiroshima, K.~Kohri, R.S.L.~Hansen, T.~Tram and S.~Hannestad, \emph{{MeV-scale reheating temperature and thermalization of oscillating neutrinos by radiative and hadronic decays of massive particles}}, \href{https://doi.org/10.1088/1475-7516/2019/12/012}{\emph{JCAP} {\bfseries 12} (2019) 012} [\href{https://arxiv.org/abs/1908.10189}{{\ttfamily 1908.10189}}].

\bibitem{Dai:2014jja}
L.~Dai, M.~Kamionkowski and J.~Wang, \emph{{Reheating constraints to inflationary models}}, \href{https://doi.org/10.1103/PhysRevLett.113.041302}{\emph{Phys. Rev. Lett.} {\bfseries 113} (2014) 041302} [\href{https://arxiv.org/abs/1404.6704}{{\ttfamily 1404.6704}}].

\bibitem{Martin:2014nya}
J.~Martin, C.~Ringeval and V.~Vennin, \emph{{Observing Inflationary Reheating}}, \href{https://doi.org/10.1103/PhysRevLett.114.081303}{\emph{Phys. Rev. Lett.} {\bfseries 114} (2015) 081303} [\href{https://arxiv.org/abs/1410.7958}{{\ttfamily 1410.7958}}].

\bibitem{Munoz:2014eqa}
J.B.~Munoz and M.~Kamionkowski, \emph{{Equation-of-State Parameter for Reheating}}, \href{https://doi.org/10.1103/PhysRevD.91.043521}{\emph{Phys. Rev. D} {\bfseries 91} (2015) 043521} [\href{https://arxiv.org/abs/1412.0656}{{\ttfamily 1412.0656}}].

\bibitem{Gong:2015qha}
J.-O.~Gong, S.~Pi and G.~Leung, \emph{{Probing reheating with primordial spectrum}}, \href{https://doi.org/10.1088/1475-7516/2015/05/027}{\emph{JCAP} {\bfseries 05} (2015) 027} [\href{https://arxiv.org/abs/1501.03604}{{\ttfamily 1501.03604}}].

\bibitem{Cook:2015vqa}
J.L.~Cook, E.~Dimastrogiovanni, D.A.~Easson and L.M.~Krauss, \emph{{Reheating predictions in single field inflation}}, \href{https://doi.org/10.1088/1475-7516/2015/04/047}{\emph{JCAP} {\bfseries 04} (2015) 047} [\href{https://arxiv.org/abs/1502.04673}{{\ttfamily 1502.04673}}].

\bibitem{Traschen:1990sw}
J.H.~Traschen and R.H.~Brandenberger, \emph{{Particle Production During Out-of-equilibrium Phase Transitions}}, \href{https://doi.org/10.1103/PhysRevD.42.2491}{\emph{Phys. Rev. D} {\bfseries 42} (1990) 2491}.

\bibitem{Kofman:1994rk}
L.~Kofman, A.D.~Linde and A.A.~Starobinsky, \emph{{Reheating after inflation}}, \href{https://doi.org/10.1103/PhysRevLett.73.3195}{\emph{Phys. Rev. Lett.} {\bfseries 73} (1994) 3195} [\href{https://arxiv.org/abs/hep-th/9405187}{{\ttfamily hep-th/9405187}}].

\bibitem{Shtanov:1994ce}
Y.~Shtanov, J.H.~Traschen and R.H.~Brandenberger, \emph{{Universe reheating after inflation}}, \href{https://doi.org/10.1103/PhysRevD.51.5438}{\emph{Phys. Rev. D} {\bfseries 51} (1995) 5438} [\href{https://arxiv.org/abs/hep-ph/9407247}{{\ttfamily hep-ph/9407247}}].

\bibitem{Kaiser:1995fb}
D.I.~Kaiser, \emph{{Post inflation reheating in an expanding universe}}, \href{https://doi.org/10.1103/PhysRevD.53.1776}{\emph{Phys. Rev. D} {\bfseries 53} (1996) 1776} [\href{https://arxiv.org/abs/astro-ph/9507108}{{\ttfamily astro-ph/9507108}}].

\bibitem{Khlebnikov:1996mc}
S.Y.~Khlebnikov and I.I.~Tkachev, \emph{{Classical decay of inflaton}}, \href{https://doi.org/10.1103/PhysRevLett.77.219}{\emph{Phys. Rev. Lett.} {\bfseries 77} (1996) 219} [\href{https://arxiv.org/abs/hep-ph/9603378}{{\ttfamily hep-ph/9603378}}].

\bibitem{Prokopec:1996rr}
T.~Prokopec and T.G.~Roos, \emph{{Lattice study of classical inflaton decay}}, \href{https://doi.org/10.1103/PhysRevD.55.3768}{\emph{Phys. Rev. D} {\bfseries 55} (1997) 3768} [\href{https://arxiv.org/abs/hep-ph/9610400}{{\ttfamily hep-ph/9610400}}].

\bibitem{Khlebnikov:1996wr}
S.Y.~Khlebnikov and I.I.~Tkachev, \emph{{The Universe after inflation: The Wide resonance case}}, \href{https://doi.org/10.1016/S0370-2693(96)01419-0}{\emph{Phys. Lett. B} {\bfseries 390} (1997) 80} [\href{https://arxiv.org/abs/hep-ph/9608458}{{\ttfamily hep-ph/9608458}}].

\bibitem{Kaiser:1997mp}
D.I.~Kaiser, \emph{{Preheating in an expanding universe: Analytic results for the massless case}}, \href{https://doi.org/10.1103/PhysRevD.56.706}{\emph{Phys. Rev. D} {\bfseries 56} (1997) 706} [\href{https://arxiv.org/abs/hep-ph/9702244}{{\ttfamily hep-ph/9702244}}].

\bibitem{Kofman:1997yn}
L.~Kofman, A.D.~Linde and A.A.~Starobinsky, \emph{{Towards the theory of reheating after inflation}}, \href{https://doi.org/10.1103/PhysRevD.56.3258}{\emph{Phys. Rev. D} {\bfseries 56} (1997) 3258} [\href{https://arxiv.org/abs/hep-ph/9704452}{{\ttfamily hep-ph/9704452}}].

\bibitem{Greene:1997fu}
P.B.~Greene, L.~Kofman, A.D.~Linde and A.A.~Starobinsky, \emph{{Structure of resonance in preheating after inflation}}, \href{https://doi.org/10.1103/PhysRevD.56.6175}{\emph{Phys. Rev. D} {\bfseries 56} (1997) 6175} [\href{https://arxiv.org/abs/hep-ph/9705347}{{\ttfamily hep-ph/9705347}}].

\bibitem{Khlebnikov:1996zt}
S.Y.~Khlebnikov and I.I.~Tkachev, \emph{{Resonant decay of Bose condensates}}, \href{https://doi.org/10.1103/PhysRevLett.79.1607}{\emph{Phys. Rev. Lett.} {\bfseries 79} (1997) 1607} [\href{https://arxiv.org/abs/hep-ph/9610477}{{\ttfamily hep-ph/9610477}}].

\bibitem{Kaiser:1997hg}
D.I.~Kaiser, \emph{{Resonance structure for preheating with massless fields}}, \href{https://doi.org/10.1103/PhysRevD.57.702}{\emph{Phys. Rev. D} {\bfseries 57} (1998) 702} [\href{https://arxiv.org/abs/hep-ph/9707516}{{\ttfamily hep-ph/9707516}}].

\bibitem{Figueroa:2020rrl}
D.G.~Figueroa, A.~Florio, F.~Torrenti and W.~Valkenburg, \emph{{The art of simulating the early Universe -- Part I}}, \href{https://doi.org/10.1088/1475-7516/2021/04/035}{\emph{JCAP} {\bfseries 04} (2021) 035} [\href{https://arxiv.org/abs/2006.15122}{{\ttfamily 2006.15122}}].

\bibitem{Bogolyubsky:1976yu}
I.L.~Bogolyubsky and V.G.~Makhankov, \emph{{Lifetime of Pulsating Solitons in Some Classical Models}}, {\emph{Pisma Zh. Eksp. Teor. Fiz.} {\bfseries 24} (1976) 15}.

\bibitem{Gleiser:1993pt}
M.~Gleiser, \emph{{Pseudostable bubbles}}, \href{https://doi.org/10.1103/PhysRevD.49.2978}{\emph{Phys. Rev. D} {\bfseries 49} (1994) 2978} [\href{https://arxiv.org/abs/hep-ph/9308279}{{\ttfamily hep-ph/9308279}}].

\bibitem{Copeland:1995fq}
E.J.~Copeland, M.~Gleiser and H.R.~Muller, \emph{{Oscillons: Resonant configurations during bubble collapse}}, \href{https://doi.org/10.1103/PhysRevD.52.1920}{\emph{Phys. Rev. D} {\bfseries 52} (1995) 1920} [\href{https://arxiv.org/abs/hep-ph/9503217}{{\ttfamily hep-ph/9503217}}].

\bibitem{Amin:2010dc}
M.A.~Amin, R.~Easther and H.~Finkel, \emph{{Inflaton Fragmentation and Oscillon Formation in Three Dimensions}}, \href{https://doi.org/10.1088/1475-7516/2010/12/001}{\emph{JCAP} {\bfseries 12} (2010) 001} [\href{https://arxiv.org/abs/1009.2505}{{\ttfamily 1009.2505}}].

\bibitem{Amin:2011hj}
M.A.~Amin, R.~Easther, H.~Finkel, R.~Flauger and M.P.~Hertzberg, \emph{{Oscillons After Inflation}}, \href{https://doi.org/10.1103/PhysRevLett.108.241302}{\emph{Phys. Rev. Lett.} {\bfseries 108} (2012) 241302} [\href{https://arxiv.org/abs/1106.3335}{{\ttfamily 1106.3335}}].

\bibitem{Antusch:2015nla}
S.~Antusch, D.~Nolde and S.~Orani, \emph{{Hill crossing during preheating after hilltop inflation}}, \href{https://doi.org/10.1088/1475-7516/2015/06/009}{\emph{JCAP} {\bfseries 06} (2015) 009} [\href{https://arxiv.org/abs/1503.06075}{{\ttfamily 1503.06075}}].

\bibitem{Lozanov:2017hjm}
K.D.~Lozanov and M.A.~Amin, \emph{{Self-resonance after inflation: oscillons, transients and radiation domination}}, \href{https://doi.org/10.1103/PhysRevD.97.023533}{\emph{Phys. Rev. D} {\bfseries 97} (2018) 023533} [\href{https://arxiv.org/abs/1710.06851}{{\ttfamily 1710.06851}}].

\bibitem{Hasegawa:2017iay}
F.~Hasegawa and J.-P.~Hong, \emph{{Inflaton fragmentation in E-models of cosmological $\alpha$-attractors}}, \href{https://doi.org/10.1103/PhysRevD.97.083514}{\emph{Phys. Rev. D} {\bfseries 97} (2018) 083514} [\href{https://arxiv.org/abs/1710.07487}{{\ttfamily 1710.07487}}].

\bibitem{Antusch:2017flz}
S.~Antusch, F.~Cefala, S.~Krippendorf, F.~Muia, S.~Orani and F.~Quevedo, \emph{{Oscillons from String Moduli}}, \href{https://doi.org/10.1007/JHEP01(2018)083}{\emph{JHEP} {\bfseries 01} (2018) 083} [\href{https://arxiv.org/abs/1708.08922}{{\ttfamily 1708.08922}}].

\bibitem{Sang:2019ndv}
Y.~Sang and Q.-G.~Huang, \emph{{Stochastic Gravitational-Wave Background from Axion-Monodromy Oscillons in String Theory During Preheating}}, \href{https://doi.org/10.1103/PhysRevD.100.063516}{\emph{Phys. Rev. D} {\bfseries 100} (2019) 063516} [\href{https://arxiv.org/abs/1905.00371}{{\ttfamily 1905.00371}}].

\bibitem{Antusch:2019qrr}
S.~Antusch, F.~Cefal\`a and F.~Torrent\'\i{}, \emph{{Properties of Oscillons in Hilltop Potentials: energies, shapes, and lifetimes}}, \href{https://doi.org/10.1088/1475-7516/2019/10/002}{\emph{JCAP} {\bfseries 10} (2019) 002} [\href{https://arxiv.org/abs/1907.00611}{{\ttfamily 1907.00611}}].

\bibitem{Ibe:2019lzv}
M.~Ibe, M.~Kawasaki, W.~Nakano and E.~Sonomoto, \emph{{Fragileness of Exact I-ball/Oscillon}}, \href{https://doi.org/10.1103/PhysRevD.100.125021}{\emph{Phys. Rev. D} {\bfseries 100} (2019) 125021} [\href{https://arxiv.org/abs/1908.11103}{{\ttfamily 1908.11103}}].

\bibitem{Kou:2019bbc}
X.-X.~Kou, C.~Tian and S.-Y.~Zhou, \emph{{Oscillon Preheating in Full General Relativity}}, \href{https://doi.org/10.1088/1361-6382/abd09f}{\emph{Class. Quant. Grav.} {\bfseries 38} (2021) 045005} [\href{https://arxiv.org/abs/1912.09658}{{\ttfamily 1912.09658}}].

\bibitem{Sang:2020kpd}
Y.~Sang and Q.-G.~Huang, \emph{{Oscillons during Dirac-Born-Infeld preheating}}, \href{https://doi.org/10.1016/j.physletb.2021.136781}{\emph{Phys. Lett. B} {\bfseries 823} (2021) 136781} [\href{https://arxiv.org/abs/2012.14697}{{\ttfamily 2012.14697}}].

\bibitem{Aurrekoetxea:2023jwd}
J.C.~Aurrekoetxea, K.~Clough and F.~Muia, \emph{{Oscillon formation during inflationary preheating with general relativity}}, \href{https://doi.org/10.1103/PhysRevD.108.023501}{\emph{Phys. Rev. D} {\bfseries 108} (2023) 023501} [\href{https://arxiv.org/abs/2304.01673}{{\ttfamily 2304.01673}}].

\bibitem{Mahbub:2023faw}
R.~Mahbub and S.S.~Mishra, \emph{{Oscillon formation from preheating in asymmetric inflationary potentials}}, \href{https://doi.org/10.1103/PhysRevD.108.063524}{\emph{Phys. Rev. D} {\bfseries 108} (2023) 063524} [\href{https://arxiv.org/abs/2303.07503}{{\ttfamily 2303.07503}}].

\bibitem{vanDissel:2023zva}
F.~van Dissel, O.~Pujolas and E.I.~Sfakianakis, \emph{{Oscillon spectroscopy}}, \href{https://doi.org/10.1007/JHEP07(2023)194}{\emph{JHEP} {\bfseries 07} (2023) 194} [\href{https://arxiv.org/abs/2303.16072}{{\ttfamily 2303.16072}}].

\bibitem{Zhou:2013tsa}
S.-Y.~Zhou, E.J.~Copeland, R.~Easther, H.~Finkel, Z.-G.~Mou and P.M.~Saffin, \emph{{Gravitational Waves from Oscillon Preheating}}, \href{https://doi.org/10.1007/JHEP10(2013)026}{\emph{JHEP} {\bfseries 10} (2013) 026} [\href{https://arxiv.org/abs/1304.6094}{{\ttfamily 1304.6094}}].

\bibitem{Antusch:2016con}
S.~Antusch, F.~Cefala and S.~Orani, \emph{{Gravitational waves from oscillons after inflation}}, \href{https://doi.org/10.1103/PhysRevLett.118.011303}{\emph{Phys. Rev. Lett.} {\bfseries 118} (2017) 011303} [\href{https://arxiv.org/abs/1607.01314}{{\ttfamily 1607.01314}}].

\bibitem{Antusch:2017vga}
S.~Antusch, F.~Cefala and S.~Orani, \emph{{What can we learn from the stochastic gravitational wave background produced by oscillons?}}, \href{https://doi.org/10.1088/1475-7516/2018/03/032}{\emph{JCAP} {\bfseries 03} (2018) 032} [\href{https://arxiv.org/abs/1712.03231}{{\ttfamily 1712.03231}}].

\bibitem{Amin:2018xfe}
M.A.~Amin, J.~Braden, E.J.~Copeland, J.T.~Giblin, C.~Solorio, Z.J.~Weiner et~al., \emph{{Gravitational waves from asymmetric oscillon dynamics?}}, \href{https://doi.org/10.1103/PhysRevD.98.024040}{\emph{Phys. Rev. D} {\bfseries 98} (2018) 024040} [\href{https://arxiv.org/abs/1803.08047}{{\ttfamily 1803.08047}}].

\bibitem{Liu:2018rrt}
J.~Liu, Z.-K.~Guo, R.-G.~Cai and G.~Shiu, \emph{{Gravitational wave production after inflation with cuspy potentials}}, \href{https://doi.org/10.1103/PhysRevD.99.103506}{\emph{Phys. Rev. D} {\bfseries 99} (2019) 103506} [\href{https://arxiv.org/abs/1812.09235}{{\ttfamily 1812.09235}}].

\bibitem{Lozanov:2019ylm}
K.D.~Lozanov and M.A.~Amin, \emph{{Gravitational perturbations from oscillons and transients after inflation}}, \href{https://doi.org/10.1103/PhysRevD.99.123504}{\emph{Phys. Rev. D} {\bfseries 99} (2019) 123504} [\href{https://arxiv.org/abs/1902.06736}{{\ttfamily 1902.06736}}].

\bibitem{Lozanov:2022yoy}
K.D.~Lozanov and V.~Takhistov, \emph{{Enhanced Gravitational Waves from Inflaton Oscillons}}, \href{https://doi.org/10.1103/PhysRevLett.130.181002}{\emph{Phys. Rev. Lett.} {\bfseries 130} (2023) 181002} [\href{https://arxiv.org/abs/2204.07152}{{\ttfamily 2204.07152}}].

\bibitem{Piani:2023aof}
M.~Piani and J.~Rubio, \emph{{Preheating in Einstein-Cartan Higgs Inflation: oscillon formation}}, \href{https://doi.org/10.1088/1475-7516/2023/12/002}{\emph{JCAP} {\bfseries 12} (2023) 002} [\href{https://arxiv.org/abs/2304.13056}{{\ttfamily 2304.13056}}].

\bibitem{Aggarwal:2020olq}
N.~Aggarwal et~al., \emph{{Challenges and opportunities of gravitational-wave searches at MHz to GHz frequencies}}, \href{https://doi.org/10.1007/s41114-021-00032-5}{\emph{Living Rev. Rel.} {\bfseries 24} (2021) 4} [\href{https://arxiv.org/abs/2011.12414}{{\ttfamily 2011.12414}}].

\bibitem{Herman:2022fau}
N.~Herman, L.~Lehoucq and A.~F\'{u}zfa, \emph{{Electromagnetic antennas for the resonant detection of the stochastic gravitational wave background}}, \href{https://doi.org/10.1103/PhysRevD.108.124009}{\emph{Phys. Rev. D} {\bfseries 108} (2023) 124009} [\href{https://arxiv.org/abs/2203.15668}{{\ttfamily 2203.15668}}].

\bibitem{Aggarwal:2025noe}
N.~Aggarwal et~al., \emph{{Challenges and Opportunities of Gravitational Wave Searches above 10 kHz}},  \href{https://arxiv.org/abs/2501.11723}{{\ttfamily 2501.11723}}.

\bibitem{Drees:2025iue}
M.~Drees and C.~Wang, \emph{{Inflaton Self Resonance, Oscillons, and Gravitational Waves in Small Field Polynomial Inflation}},  \href{https://arxiv.org/abs/2501.13811}{{\ttfamily 2501.13811}}.

\bibitem{Antusch:2015ziz}
S.~Antusch and S.~Orani, \emph{{Impact of other scalar fields on oscillons after hilltop inflation}}, \href{https://doi.org/10.1088/1475-7516/2016/03/026}{\emph{JCAP} {\bfseries 03} (2016) 026} [\href{https://arxiv.org/abs/1511.02336}{{\ttfamily 1511.02336}}].

\bibitem{Shafi:2024jig}
M.~Shafi, E.J.~Copeland, R.~Mahbub, S.S.~Mishra and S.~Basak, \emph{{Formation and decay of oscillons after inflation in the presence of an external coupling. Part I. Lattice simulations}}, \href{https://doi.org/10.1088/1475-7516/2024/10/082}{\emph{JCAP} {\bfseries 10} (2024) 082} [\href{https://arxiv.org/abs/2406.00108}{{\ttfamily 2406.00108}}].

\bibitem{Utiyama:1956sy}
R.~Utiyama, \emph{{Invariant theoretical interpretation of interaction}}, \href{https://doi.org/10.1103/PhysRev.101.1597}{\emph{Phys. Rev.} {\bfseries 101} (1956) 1597}.

\bibitem{Kibble:1961ba}
T.W.B.~Kibble, \emph{{Lorentz invariance and the gravitational field}}, \href{https://doi.org/10.1063/1.1703702}{\emph{J. Math. Phys.} {\bfseries 2} (1961) 212}.

\bibitem{Ferraris1982}
M.~Ferraris, M.~Francaviglia and C.~Reina, \emph{{Variational formulation of general relativity from 1915 to 1925 Palatini's method discovered by Einstein in 1925}}, {\emph{General Relativity and Gravitation} {\bfseries 14} (1982) 243}.

\bibitem{Shaposhnikov:2020aen}
M.~Shaposhnikov, A.~Shkerin, I.~Timiryasov and S.~Zell, \emph{{Einstein-Cartan Portal to Dark Matter}}, \href{https://doi.org/10.1103/PhysRevLett.127.169901}{\emph{Phys. Rev. Lett.} {\bfseries 126} (2021) 161301} [\href{https://arxiv.org/abs/2008.11686}{{\ttfamily 2008.11686}}].

\bibitem{Karananas:2021zkl}
G.K.~Karananas, M.~Shaposhnikov, A.~Shkerin and S.~Zell, \emph{{Matter matters in Einstein-Cartan gravity}}, \href{https://doi.org/10.1103/PhysRevD.104.064036}{\emph{Phys. Rev. D} {\bfseries 104} (2021) 064036} [\href{https://arxiv.org/abs/2106.13811}{{\ttfamily 2106.13811}}].

\bibitem{Barker:2024dhb}
W.~Barker and S.~Zell, \emph{{Consistent particle physics in metric-affine gravity from extended projective symmetry}},  \href{https://arxiv.org/abs/2402.14917}{{\ttfamily 2402.14917}}.

\bibitem{Langvik:2020nrs}
M.~L\r{a}ngvik, J.-M.~Ojanper\"a, S.~Raatikainen and S.~R\"as\"anen, \emph{{Higgs inflation with the Holst and the Nieh\textendash{}Yan term}}, \href{https://doi.org/10.1103/PhysRevD.103.083514}{\emph{Phys. Rev. D} {\bfseries 103} (2021) 083514} [\href{https://arxiv.org/abs/2007.12595}{{\ttfamily 2007.12595}}].

\bibitem{Shaposhnikov:2020gts}
M.~Shaposhnikov, A.~Shkerin, I.~Timiryasov and S.~Zell, \emph{{Higgs inflation in Einstein-Cartan gravity}}, \href{https://doi.org/10.1088/1475-7516/2021/02/008}{\emph{JCAP} {\bfseries 02} (2021) 008} [\href{https://arxiv.org/abs/2007.14978}{{\ttfamily 2007.14978}}].

\bibitem{Rigouzzo:2022yan}
C.~Rigouzzo and S.~Zell, \emph{{Coupling metric-affine gravity to a Higgs-like scalar field}}, \href{https://doi.org/10.1103/PhysRevD.106.024015}{\emph{Phys. Rev. D} {\bfseries 106} (2022) 024015} [\href{https://arxiv.org/abs/2204.03003}{{\ttfamily 2204.03003}}].

\bibitem{Gialamas:2024iyu}
I.D.~Gialamas and K.~Tamvakis, \emph{{Inflation in Weyl-invariant Einstein-Cartan gravity}}, \href{https://doi.org/10.1103/PhysRevD.111.044007}{\emph{Phys. Rev. D} {\bfseries 111} (2025) 044007} [\href{https://arxiv.org/abs/2410.16364}{{\ttfamily 2410.16364}}].

\bibitem{Rigouzzo:2023sbb}
C.~Rigouzzo and S.~Zell, \emph{{Coupling metric-affine gravity to the standard model and dark matter fermions}}, \href{https://doi.org/10.1103/PhysRevD.108.124067}{\emph{Phys. Rev. D} {\bfseries 108} (2023) 124067} [\href{https://arxiv.org/abs/2306.13134}{{\ttfamily 2306.13134}}].

\bibitem{Piani:2022gon}
M.~Piani and J.~Rubio, \emph{{Higgs-Dilaton inflation in Einstein-Cartan gravity}}, \href{https://doi.org/10.1088/1475-7516/2022/05/009}{\emph{JCAP} {\bfseries 05} (2022) 009} [\href{https://arxiv.org/abs/2202.04665}{{\ttfamily 2202.04665}}].

\bibitem{Nieh:1981ww}
H.T.~Nieh and M.L.~Yan, \emph{{An Identity in Riemann-cartan Geometry}}, \href{https://doi.org/10.1063/1.525379}{\emph{J. Math. Phys.} {\bfseries 23} (1982) 373}.

\bibitem{Nieh:2008btw}
H.T.~Nieh, \emph{{A Torsional Topological Invariant}},  in \emph{{Conference in Honor of C.N. Yang's 85th Birthday}: {Statistical Physics, High Energy, Condensed Matter and Mathematical Physics}}, pp.~29--37, 2008, \href{https://doi.org/10.1142/9789812794185_0003}{DOI} [\href{https://arxiv.org/abs/1309.0915}{{\ttfamily 1309.0915}}].

\bibitem{Bezrukov:2007ep}
F.L.~Bezrukov and M.~Shaposhnikov, \emph{{The Standard Model Higgs boson as the inflaton}}, \href{https://doi.org/10.1016/j.physletb.2007.11.072}{\emph{Phys. Lett. B} {\bfseries 659} (2008) 703} [\href{https://arxiv.org/abs/0710.3755}{{\ttfamily 0710.3755}}].

\bibitem{Barbon:2009ya}
J.L.F.~Barbon and J.R.~Espinosa, \emph{{On the Naturalness of Higgs Inflation}}, \href{https://doi.org/10.1103/PhysRevD.79.081302}{\emph{Phys. Rev. D} {\bfseries 79} (2009) 081302} [\href{https://arxiv.org/abs/0903.0355}{{\ttfamily 0903.0355}}].

\bibitem{Bezrukov:2009db}
F.~Bezrukov and M.~Shaposhnikov, \emph{{Standard Model Higgs boson mass from inflation: Two loop analysis}}, \href{https://doi.org/10.1088/1126-6708/2009/07/089}{\emph{JHEP} {\bfseries 07} (2009) 089} [\href{https://arxiv.org/abs/0904.1537}{{\ttfamily 0904.1537}}].

\bibitem{Burgess:2010zq}
C.P.~Burgess, H.M.~Lee and M.~Trott, \emph{{Comment on Higgs Inflation and Naturalness}}, \href{https://doi.org/10.1007/JHEP07(2010)007}{\emph{JHEP} {\bfseries 07} (2010) 007} [\href{https://arxiv.org/abs/1002.2730}{{\ttfamily 1002.2730}}].

\bibitem{Bezrukov:2010jz}
F.~Bezrukov, A.~Magnin, M.~Shaposhnikov and S.~Sibiryakov, \emph{{Higgs inflation: consistency and generalisations}}, \href{https://doi.org/10.1007/JHEP01(2011)016}{\emph{JHEP} {\bfseries 01} (2011) 016} [\href{https://arxiv.org/abs/1008.5157}{{\ttfamily 1008.5157}}].

\bibitem{Giudice:2010ka}
G.F.~Giudice and H.M.~Lee, \emph{{Unitarizing Higgs Inflation}}, \href{https://doi.org/10.1016/j.physletb.2010.10.035}{\emph{Phys. Lett. B} {\bfseries 694} (2011) 294} [\href{https://arxiv.org/abs/1010.1417}{{\ttfamily 1010.1417}}].

\bibitem{Bezrukov:2014bra}
F.~Bezrukov and M.~Shaposhnikov, \emph{{Higgs inflation at the critical point}}, \href{https://doi.org/10.1016/j.physletb.2014.05.074}{\emph{Phys. Lett. B} {\bfseries 734} (2014) 249} [\href{https://arxiv.org/abs/1403.6078}{{\ttfamily 1403.6078}}].

\bibitem{Hamada:2014iga}
Y.~Hamada, H.~Kawai, K.-y.~Oda and S.C.~Park, \emph{{Higgs Inflation is Still Alive after the Results from BICEP2}}, \href{https://doi.org/10.1103/PhysRevLett.112.241301}{\emph{Phys. Rev. Lett.} {\bfseries 112} (2014) 241301} [\href{https://arxiv.org/abs/1403.5043}{{\ttfamily 1403.5043}}].

\bibitem{Hamada:2014wna}
Y.~Hamada, H.~Kawai, K.-y.~Oda and S.C.~Park, \emph{{Higgs inflation from Standard Model criticality}}, \href{https://doi.org/10.1103/PhysRevD.91.053008}{\emph{Phys. Rev. D} {\bfseries 91} (2015) 053008} [\href{https://arxiv.org/abs/1408.4864}{{\ttfamily 1408.4864}}].

\bibitem{George:2015nza}
D.P.~George, S.~Mooij and M.~Postma, \emph{{Quantum corrections in Higgs inflation: the Standard Model case}}, \href{https://doi.org/10.1088/1475-7516/2016/04/006}{\emph{JCAP} {\bfseries 04} (2016) 006} [\href{https://arxiv.org/abs/1508.04660}{{\ttfamily 1508.04660}}].

\bibitem{Fumagalli:2016lls}
J.~Fumagalli and M.~Postma, \emph{{UV (in)sensitivity of Higgs inflation}}, \href{https://doi.org/10.1007/JHEP05(2016)049}{\emph{JHEP} {\bfseries 05} (2016) 049} [\href{https://arxiv.org/abs/1602.07234}{{\ttfamily 1602.07234}}].

\bibitem{Bezrukov:2017dyv}
F.~Bezrukov, M.~Pauly and J.~Rubio, \emph{{On the robustness of the primordial power spectrum in renormalized Higgs inflation}}, \href{https://doi.org/10.1088/1475-7516/2018/02/040}{\emph{JCAP} {\bfseries 02} (2018) 040} [\href{https://arxiv.org/abs/1706.05007}{{\ttfamily 1706.05007}}].

\bibitem{Shaposhnikov:2008xb}
M.~Shaposhnikov and D.~Zenhausern, \emph{{Scale invariance, unimodular gravity and dark energy}}, \href{https://doi.org/10.1016/j.physletb.2008.11.054}{\emph{Phys. Lett. B} {\bfseries 671} (2009) 187} [\href{https://arxiv.org/abs/0809.3395}{{\ttfamily 0809.3395}}].

\bibitem{Garcia-Bellido:2011kqb}
J.~Garcia-Bellido, J.~Rubio, M.~Shaposhnikov and D.~Zenhausern, \emph{{Higgs-Dilaton Cosmology: From the Early to the Late Universe}}, \href{https://doi.org/10.1103/PhysRevD.84.123504}{\emph{Phys. Rev. D} {\bfseries 84} (2011) 123504} [\href{https://arxiv.org/abs/1107.2163}{{\ttfamily 1107.2163}}].

\bibitem{Garcia-Bellido:2012npk}
J.~Garcia-Bellido, J.~Rubio and M.~Shaposhnikov, \emph{{Higgs-Dilaton cosmology: Are there extra relativistic species?}}, \href{https://doi.org/10.1016/j.physletb.2012.10.075}{\emph{Phys. Lett. B} {\bfseries 718} (2012) 507} [\href{https://arxiv.org/abs/1209.2119}{{\ttfamily 1209.2119}}].

\bibitem{Bezrukov:2012hx}
F.~Bezrukov, G.K.~Karananas, J.~Rubio and M.~Shaposhnikov, \emph{{Higgs-Dilaton Cosmology: an effective field theory approach}}, \href{https://doi.org/10.1103/PhysRevD.87.096001}{\emph{Phys. Rev. D} {\bfseries 87} (2013) 096001} [\href{https://arxiv.org/abs/1212.4148}{{\ttfamily 1212.4148}}].

\bibitem{Rubio:2014wta}
J.~Rubio and M.~Shaposhnikov, \emph{{Higgs-Dilaton cosmology: Universality versus criticality}}, \href{https://doi.org/10.1103/PhysRevD.90.027307}{\emph{Phys. Rev. D} {\bfseries 90} (2014) 027307} [\href{https://arxiv.org/abs/1406.5182}{{\ttfamily 1406.5182}}].

\bibitem{Karananas:2016kyt}
G.K.~Karananas and J.~Rubio, \emph{{On the geometrical interpretation of scale-invariant models of inflation}}, \href{https://doi.org/10.1016/j.physletb.2016.08.037}{\emph{Phys. Lett. B} {\bfseries 761} (2016) 223} [\href{https://arxiv.org/abs/1606.08848}{{\ttfamily 1606.08848}}].

\bibitem{Trashorras:2016azl}
M.~Trashorras, S.~Nesseris and J.~Garcia-Bellido, \emph{{Cosmological Constraints on Higgs-Dilaton Inflation}}, \href{https://doi.org/10.1103/PhysRevD.94.063511}{\emph{Phys. Rev. D} {\bfseries 94} (2016) 063511} [\href{https://arxiv.org/abs/1604.06760}{{\ttfamily 1604.06760}}].

\bibitem{Casas:2017wjh}
S.~Casas, M.~Pauly and J.~Rubio, \emph{{Higgs-dilaton cosmology: An inflation\textendash{}dark-energy connection and forecasts for future galaxy surveys}}, \href{https://doi.org/10.1103/PhysRevD.97.043520}{\emph{Phys. Rev. D} {\bfseries 97} (2018) 043520} [\href{https://arxiv.org/abs/1712.04956}{{\ttfamily 1712.04956}}].

\bibitem{Tokareva:2017nng}
A.~Tokareva, \emph{{A minimal scale invariant axion solution to the strong CP-problem}}, \href{https://doi.org/10.1140/epjc/s10052-018-5883-0}{\emph{Eur. Phys. J. C} {\bfseries 78} (2018) 423} [\href{https://arxiv.org/abs/1705.10836}{{\ttfamily 1705.10836}}].

\bibitem{Casas:2018fum}
S.~Casas, G.K.~Karananas, M.~Pauly and J.~Rubio, \emph{{Scale-invariant alternatives to general relativity. III. The inflation-dark energy connection}}, \href{https://doi.org/10.1103/PhysRevD.99.063512}{\emph{Phys. Rev. D} {\bfseries 99} (2019) 063512} [\href{https://arxiv.org/abs/1811.05984}{{\ttfamily 1811.05984}}].

\bibitem{Shaposhnikov:2018jag}
M.~Shaposhnikov and A.~Shkerin, \emph{{Gravity, Scale Invariance and the Hierarchy Problem}}, \href{https://doi.org/10.1007/JHEP10(2018)024}{\emph{JHEP} {\bfseries 10} (2018) 024} [\href{https://arxiv.org/abs/1804.06376}{{\ttfamily 1804.06376}}].

\bibitem{Herrero-Valea:2019hde}
M.~Herrero-Valea, I.~Timiryasov and A.~Tokareva, \emph{{To Positivity and Beyond, where Higgs-Dilaton Inflation has never gone before}}, \href{https://doi.org/10.1088/1475-7516/2019/11/042}{\emph{JCAP} {\bfseries 11} (2019) 042} [\href{https://arxiv.org/abs/1905.08816}{{\ttfamily 1905.08816}}].

\bibitem{Karananas:2020qkp}
G.K.~Karananas, M.~Michel and J.~Rubio, \emph{{One residue to rule them all: Electroweak symmetry breaking, inflation and field-space geometry}}, \href{https://doi.org/10.1016/j.physletb.2020.135876}{\emph{Phys. Lett. B} {\bfseries 811} (2020) 135876} [\href{https://arxiv.org/abs/2006.11290}{{\ttfamily 2006.11290}}].

\bibitem{Rubio:2020zht}
J.~Rubio, \emph{{Scale symmetry, the Higgs and the Cosmos}}, \href{https://doi.org/10.22323/1.376.0074}{\emph{PoS} {\bfseries CORFU2019} (2020) 074} [\href{https://arxiv.org/abs/2004.00039}{{\ttfamily 2004.00039}}].

\bibitem{Shaposhnikov:2020frq}
M.~Shaposhnikov, A.~Shkerin, I.~Timiryasov and S.~Zell, \emph{{Einstein-Cartan gravity, matter, and scale-invariant generalization~}}, \href{https://doi.org/10.1007/JHEP08(2021)162}{\emph{JHEP} {\bfseries 10} (2020) 177} [\href{https://arxiv.org/abs/2007.16158}{{\ttfamily 2007.16158}}].

\bibitem{Karananas:2021gco}
G.K.~Karananas, M.~Shaposhnikov, A.~Shkerin and S.~Zell, \emph{{Scale and Weyl invariance in Einstein-Cartan gravity}}, \href{https://doi.org/10.1103/PhysRevD.104.124014}{\emph{Phys. Rev. D} {\bfseries 104} (2021) 124014} [\href{https://arxiv.org/abs/2108.05897}{{\ttfamily 2108.05897}}].

\bibitem{Gialamas:2021enw}
I.D.~Gialamas, A.~Karam, T.D.~Pappas and V.C.~Spanos, \emph{{Scale-invariant quadratic gravity and inflation in the Palatini formalism}}, \href{https://doi.org/10.1103/PhysRevD.104.023521}{\emph{Phys. Rev. D} {\bfseries 104} (2021) 023521} [\href{https://arxiv.org/abs/2104.04550}{{\ttfamily 2104.04550}}].

\bibitem{Belokon:2022pqf}
A.I.~Belokon and A.~Tokareva, \emph{{Higgs-dilaton model revisited: can dilaton act as QCD axion?}},  \href{https://arxiv.org/abs/2212.06739}{{\ttfamily 2212.06739}}.

\bibitem{Karananas:2023zgg}
G.K.~Karananas, M.~Shaposhnikov and S.~Zell, \emph{{Scale invariant Einstein-Cartan gravity and flat space conformal symmetry}}, \href{https://doi.org/10.1007/JHEP11(2023)171}{\emph{JHEP} {\bfseries 11} (2023) 171} [\href{https://arxiv.org/abs/2307.11151}{{\ttfamily 2307.11151}}].

\bibitem{Rubio:2018ogq}
J.~Rubio, \emph{{Higgs inflation}}, \href{https://doi.org/10.3389/fspas.2018.00050}{\emph{Front. Astron. Space Sci.} {\bfseries 5} (2019) 50} [\href{https://arxiv.org/abs/1807.02376}{{\ttfamily 1807.02376}}].

\bibitem{Bauer:2008zj}
F.~Bauer and D.A.~Demir, \emph{{Inflation with Non-Minimal Coupling: Metric versus Palatini Formulations}}, \href{https://doi.org/10.1016/j.physletb.2008.06.014}{\emph{Phys. Lett. B} {\bfseries 665} (2008) 222} [\href{https://arxiv.org/abs/0803.2664}{{\ttfamily 0803.2664}}].

\bibitem{Bauer:2010jg}
F.~Bauer and D.A.~Demir, \emph{{Higgs-Palatini Inflation and Unitarity}}, \href{https://doi.org/10.1016/j.physletb.2011.03.042}{\emph{Phys. Lett. B} {\bfseries 698} (2011) 425} [\href{https://arxiv.org/abs/1012.2900}{{\ttfamily 1012.2900}}].

\bibitem{Rasanen:2017ivk}
S.~Rasanen and P.~Wahlman, \emph{{Higgs inflation with loop corrections in the Palatini formulation}}, \href{https://doi.org/10.1088/1475-7516/2017/11/047}{\emph{JCAP} {\bfseries 11} (2017) 047} [\href{https://arxiv.org/abs/1709.07853}{{\ttfamily 1709.07853}}].

\bibitem{Enckell:2018kkc}
V.-M.~Enckell, K.~Enqvist, S.~Rasanen and E.~Tomberg, \emph{{Higgs inflation at the hilltop}}, \href{https://doi.org/10.1088/1475-7516/2018/06/005}{\emph{JCAP} {\bfseries 06} (2018) 005} [\href{https://arxiv.org/abs/1802.09299}{{\ttfamily 1802.09299}}].

\bibitem{Rasanen:2018fom}
S.~Rasanen and E.~Tomberg, \emph{{Planck scale black hole dark matter from Higgs inflation}}, \href{https://doi.org/10.1088/1475-7516/2019/01/038}{\emph{JCAP} {\bfseries 01} (2019) 038} [\href{https://arxiv.org/abs/1810.12608}{{\ttfamily 1810.12608}}].

\bibitem{Rasanen:2018ihz}
S.~Rasanen, \emph{{Higgs inflation in the Palatini formulation with kinetic terms for the metric}}, \href{https://doi.org/10.21105/astro.1811.09514}{\emph{Open J. Astrophys.} {\bfseries 2} (2019) 1} [\href{https://arxiv.org/abs/1811.09514}{{\ttfamily 1811.09514}}].

\bibitem{Gialamas:2019nly}
I.D.~Gialamas and A.B.~Lahanas, \emph{{Reheating in $R^2$ Palatini inflationary models}}, \href{https://doi.org/10.1103/PhysRevD.101.084007}{\emph{Phys. Rev. D} {\bfseries 101} (2020) 084007} [\href{https://arxiv.org/abs/1911.11513}{{\ttfamily 1911.11513}}].

\bibitem{Rubio:2019ypq}
J.~Rubio and E.S.~Tomberg, \emph{{Preheating in Palatini Higgs inflation}}, \href{https://doi.org/10.1088/1475-7516/2019/04/021}{\emph{JCAP} {\bfseries 04} (2019) 021} [\href{https://arxiv.org/abs/1902.10148}{{\ttfamily 1902.10148}}].

\bibitem{Shaposhnikov:2020fdv}
M.~Shaposhnikov, A.~Shkerin and S.~Zell, \emph{{Quantum Effects in Palatini Higgs Inflation}}, \href{https://doi.org/10.1088/1475-7516/2020/07/064}{\emph{JCAP} {\bfseries 07} (2020) 064} [\href{https://arxiv.org/abs/2002.07105}{{\ttfamily 2002.07105}}].

\bibitem{Annala:2021zdt}
J.~Annala and S.~Rasanen, \emph{{Inflation with R (\ensuremath{\alpha}\ensuremath{\beta}) terms in the Palatini formulation}}, \href{https://doi.org/10.1088/1475-7516/2021/09/032}{\emph{JCAP} {\bfseries 09} (2021) 032} [\href{https://arxiv.org/abs/2106.12422}{{\ttfamily 2106.12422}}].

\bibitem{Dux:2022kuk}
F.~Dux, A.~Florio, J.~Klari\'c, A.~Shkerin and I.~Timiryasov, \emph{{Preheating in Palatini Higgs inflation on the lattice}}, \href{https://doi.org/10.1088/1475-7516/2022/09/015}{\emph{JCAP} {\bfseries 09} (2022) 015} [\href{https://arxiv.org/abs/2203.13286}{{\ttfamily 2203.13286}}].

\bibitem{Poisson:2023tja}
A.~Poisson, I.~Timiryasov and S.~Zell, \emph{{Critical points in Palatini Higgs inflation with small non-minimal coupling}}, \href{https://doi.org/10.1007/JHEP03(2024)130}{\emph{JHEP} {\bfseries 03} (2024) 130} [\href{https://arxiv.org/abs/2306.03893}{{\ttfamily 2306.03893}}].

\bibitem{Turner:1983he}
M.S.~Turner, \emph{{Coherent Scalar Field Oscillations in an Expanding Universe}}, \href{https://doi.org/10.1103/PhysRevD.28.1243}{\emph{Phys. Rev. D} {\bfseries 28} (1983) 1243}.

\bibitem{Lozanov:2016hid}
K.D.~Lozanov and M.A.~Amin, \emph{{Equation of State and Duration to Radiation Domination after Inflation}}, \href{https://doi.org/10.1103/PhysRevLett.119.061301}{\emph{Phys. Rev. Lett.} {\bfseries 119} (2017) 061301} [\href{https://arxiv.org/abs/1608.01213}{{\ttfamily 1608.01213}}].

\bibitem{Garcia-Bellido:2008ycs}
J.~Garcia-Bellido, D.G.~Figueroa and J.~Rubio, \emph{{Preheating in the Standard Model with the Higgs-Inflaton coupled to gravity}}, \href{https://doi.org/10.1103/PhysRevD.79.063531}{\emph{Phys. Rev. D} {\bfseries 79} (2009) 063531} [\href{https://arxiv.org/abs/0812.4624}{{\ttfamily 0812.4624}}].

\bibitem{Bezrukov:2008ut}
F.~Bezrukov, D.~Gorbunov and M.~Shaposhnikov, \emph{{On initial conditions for the Hot Big Bang}}, \href{https://doi.org/10.1088/1475-7516/2009/06/029}{\emph{JCAP} {\bfseries 06} (2009) 029} [\href{https://arxiv.org/abs/0812.3622}{{\ttfamily 0812.3622}}].

\bibitem{Bezrukov:2014ipa}
F.~Bezrukov, J.~Rubio and M.~Shaposhnikov, \emph{{Living beyond the edge: Higgs inflation and vacuum metastability}}, \href{https://doi.org/10.1103/PhysRevD.92.083512}{\emph{Phys. Rev. D} {\bfseries 92} (2015) 083512} [\href{https://arxiv.org/abs/1412.3811}{{\ttfamily 1412.3811}}].

\bibitem{Rubio:2015zia}
J.~Rubio, \emph{{Higgs inflation and vacuum stability}}, \href{https://doi.org/10.1088/1742-6596/631/1/012032}{\emph{J. Phys. Conf. Ser.} {\bfseries 631} (2015) 012032} [\href{https://arxiv.org/abs/1502.07952}{{\ttfamily 1502.07952}}].

\bibitem{Repond:2016sol}
J.~Repond and J.~Rubio, \emph{{Combined Preheating on the lattice with applications to Higgs inflation}}, \href{https://doi.org/10.1088/1475-7516/2016/07/043}{\emph{JCAP} {\bfseries 07} (2016) 043} [\href{https://arxiv.org/abs/1604.08238}{{\ttfamily 1604.08238}}].

\bibitem{DeCross:2015uza}
M.P.~DeCross, D.I.~Kaiser, A.~Prabhu, C.~Prescod-Weinstein and E.I.~Sfakianakis, \emph{{Preheating after Multifield Inflation with Nonminimal Couplings, I: Covariant Formalism and Attractor Behavior}}, \href{https://doi.org/10.1103/PhysRevD.97.023526}{\emph{Phys. Rev. D} {\bfseries 97} (2018) 023526} [\href{https://arxiv.org/abs/1510.08553}{{\ttfamily 1510.08553}}].

\bibitem{DeCross:2016fdz}
M.P.~DeCross, D.I.~Kaiser, A.~Prabhu, C.~Prescod-Weinstein and E.I.~Sfakianakis, \emph{{Preheating after multifield inflation with nonminimal couplings, II: Resonance Structure}}, \href{https://doi.org/10.1103/PhysRevD.97.023527}{\emph{Phys. Rev. D} {\bfseries 97} (2018) 023527} [\href{https://arxiv.org/abs/1610.08868}{{\ttfamily 1610.08868}}].

\bibitem{DeCross:2016cbs}
M.P.~DeCross, D.I.~Kaiser, A.~Prabhu, C.~Prescod-Weinstein and E.I.~Sfakianakis, \emph{{Preheating after multifield inflation with nonminimal couplings, III: Dynamical spacetime results}}, \href{https://doi.org/10.1103/PhysRevD.97.023528}{\emph{Phys. Rev. D} {\bfseries 97} (2018) 023528} [\href{https://arxiv.org/abs/1610.08916}{{\ttfamily 1610.08916}}].

\bibitem{Ema:2016dny}
Y.~Ema, R.~Jinno, K.~Mukaida and K.~Nakayama, \emph{{Violent Preheating in Inflation with Nonminimal Coupling}}, \href{https://doi.org/10.1088/1475-7516/2017/02/045}{\emph{JCAP} {\bfseries 02} (2017) 045} [\href{https://arxiv.org/abs/1609.05209}{{\ttfamily 1609.05209}}].

\bibitem{Sfakianakis:2018lzf}
E.I.~Sfakianakis and J.~van~de Vis, \emph{{Preheating after Higgs Inflation: Self-Resonance and Gauge boson production}}, \href{https://doi.org/10.1103/PhysRevD.99.083519}{\emph{Phys. Rev. D} {\bfseries 99} (2019) 083519} [\href{https://arxiv.org/abs/1810.01304}{{\ttfamily 1810.01304}}].

\bibitem{Figueroa:2021yhd}
D.G.~Figueroa, A.~Florio, F.~Torrenti and W.~Valkenburg, \emph{{CosmoLattice}},  \href{https://arxiv.org/abs/2102.01031}{{\ttfamily 2102.01031}}.

\bibitem{Figueroa:2023xmq}
D.G.~Figueroa, A.~Florio and F.~Torrenti, \emph{{Present and future of ${\mathcal{C}}$ osmo ${\mathcal{L}}$ attice}}, \href{https://doi.org/10.1088/1361-6633/ad616a}{\emph{Rept. Prog. Phys.} {\bfseries 87} (2024) 094901} [\href{https://arxiv.org/abs/2312.15056}{{\ttfamily 2312.15056}}].

\bibitem{Zhou:2024mea}
S.-Y.~Zhou, \emph{{Non-topological solitons and quasi-solitons}}, \href{https://doi.org/10.1088/1361-6633/adc69e}{\emph{Rept. Prog. Phys.} {\bfseries 88} (2025) 046901} [\href{https://arxiv.org/abs/2411.16604}{{\ttfamily 2411.16604}}].

\bibitem{Allahverdi:2020bys}
R.~Allahverdi et~al., \emph{{The First Three Seconds: a Review of Possible Expansion Histories of the Early Universe}}, \href{https://doi.org/10.21105/astro.2006.16182}{\emph{Open J. Astrophys.} {\bfseries 4} (2021) astro.2006.16182} [\href{https://arxiv.org/abs/2006.16182}{{\ttfamily 2006.16182}}].

\bibitem{Zhang:2020bec}
H.-Y.~Zhang, M.A.~Amin, E.J.~Copeland, P.M.~Saffin and K.D.~Lozanov, \emph{{Classical Decay Rates of Oscillons}}, \href{https://doi.org/10.1088/1475-7516/2020/07/055}{\emph{JCAP} {\bfseries 07} (2020) 055} [\href{https://arxiv.org/abs/2004.01202}{{\ttfamily 2004.01202}}].

\bibitem{Antusch:2021aiw}
S.~Antusch, D.G.~Figueroa, K.~Marschall and F.~Torrenti, \emph{{Characterizing the postinflationary reheating history: Single daughter field with quadratic-quadratic interaction}}, \href{https://doi.org/10.1103/PhysRevD.105.043532}{\emph{Phys. Rev. D} {\bfseries 105} (2022) 043532} [\href{https://arxiv.org/abs/2112.11280}{{\ttfamily 2112.11280}}].

\bibitem{LiteBIRD:2022cnt}
{\scshape LiteBIRD} collaboration, \emph{{Probing Cosmic Inflation with the LiteBIRD Cosmic Microwave Background Polarization Survey}}, \href{https://doi.org/10.1093/ptep/ptac150}{\emph{PTEP} {\bfseries 2023} (2023) 042F01} [\href{https://arxiv.org/abs/2202.02773}{{\ttfamily 2202.02773}}].

\bibitem{tech}
J.~Baeza-Ballesteros, D.G.~Figueroa, A.~Florio and N.~Loayza, ``{CosmoLattice Technical Note II: Gravitational waves}.'' \url{https://cosmolattice.net/assets/technical_notes/CosmoLattice_TechnicalNote_GWs.pdf}, 2022.

\bibitem{Dufaux:2007pt}
J.F.~Dufaux, A.~Bergman, G.N.~Felder, L.~Kofman and J.-P.~Uzan, \emph{{Theory and Numerics of Gravitational Waves from Preheating after Inflation}}, \href{https://doi.org/10.1103/PhysRevD.76.123517}{\emph{Phys. Rev. D} {\bfseries 76} (2007) 123517} [\href{https://arxiv.org/abs/0707.0875}{{\ttfamily 0707.0875}}].

\bibitem{Hertzberg:2010yz}
M.P.~Hertzberg, \emph{{Quantum Radiation of Oscillons}}, \href{https://doi.org/10.1103/PhysRevD.82.045022}{\emph{Phys. Rev. D} {\bfseries 82} (2010) 045022} [\href{https://arxiv.org/abs/1003.3459}{{\ttfamily 1003.3459}}].

\bibitem{Cotner:2018vug}
E.~Cotner, A.~Kusenko and V.~Takhistov, \emph{{Primordial Black Holes from Inflaton Fragmentation into Oscillons}}, \href{https://doi.org/10.1103/PhysRevD.98.083513}{\emph{Phys. Rev. D} {\bfseries 98} (2018) 083513} [\href{https://arxiv.org/abs/1801.03321}{{\ttfamily 1801.03321}}].

\bibitem{Amin:2019ums}
M.A.~Amin and P.~Mocz, \emph{{Formation, gravitational clustering, and interactions of nonrelativistic solitons in an expanding universe}}, \href{https://doi.org/10.1103/PhysRevD.100.063507}{\emph{Phys. Rev. D} {\bfseries 100} (2019) 063507} [\href{https://arxiv.org/abs/1902.07261}{{\ttfamily 1902.07261}}].

\bibitem{Lozanov:2023aez}
K.D.~Lozanov, M.~Sasaki and V.~Takhistov, \emph{{Universal gravitational wave signatures of cosmological solitons}}, \href{https://doi.org/10.1088/1475-7516/2025/01/094}{\emph{JCAP} {\bfseries 01} (2025) 094} [\href{https://arxiv.org/abs/2304.06709}{{\ttfamily 2304.06709}}].

\bibitem{Lozanov:2023knf}
K.D.~Lozanov, M.~Sasaki and V.~Takhistov, \emph{{Universal gravitational waves from interacting and clustered solitons}}, \href{https://doi.org/10.1016/j.physletb.2023.138392}{\emph{Phys. Lett. B} {\bfseries 848} (2024) 138392} [\href{https://arxiv.org/abs/2309.14193}{{\ttfamily 2309.14193}}].

\bibitem{Kallosh:2013hoa}
R.~Kallosh and A.~Linde, \emph{{Universality Class in Conformal Inflation}}, \href{https://doi.org/10.1088/1475-7516/2013/07/002}{\emph{JCAP} {\bfseries 07} (2013) 002} [\href{https://arxiv.org/abs/1306.5220}{{\ttfamily 1306.5220}}].

\bibitem{Kallosh:2013yoa}
R.~Kallosh, A.~Linde and D.~Roest, \emph{{Superconformal Inflationary $\alpha$-Attractors}}, \href{https://doi.org/10.1007/JHEP11(2013)198}{\emph{JHEP} {\bfseries 11} (2013) 198} [\href{https://arxiv.org/abs/1311.0472}{{\ttfamily 1311.0472}}].

\bibitem{Galante:2014ifa}
M.~Galante, R.~Kallosh, A.~Linde and D.~Roest, \emph{{Unity of Cosmological Inflation Attractors}}, \href{https://doi.org/10.1103/PhysRevLett.114.141302}{\emph{Phys. Rev. Lett.} {\bfseries 114} (2015) 141302} [\href{https://arxiv.org/abs/1412.3797}{{\ttfamily 1412.3797}}].

\bibitem{Carrasco:2015pla}
J.J.M.~Carrasco, R.~Kallosh and A.~Linde, \emph{{$\alpha $-Attractors: Planck, LHC and Dark Energy}}, \href{https://doi.org/10.1007/JHEP10(2015)147}{\emph{JHEP} {\bfseries 10} (2015) 147} [\href{https://arxiv.org/abs/1506.01708}{{\ttfamily 1506.01708}}].

\bibitem{Carrasco:2015rva}
J.J.M.~Carrasco, R.~Kallosh and A.~Linde, \emph{{Cosmological Attractors and Initial Conditions for Inflation}}, \href{https://doi.org/10.1103/PhysRevD.92.063519}{\emph{Phys. Rev. D} {\bfseries 92} (2015) 063519} [\href{https://arxiv.org/abs/1506.00936}{{\ttfamily 1506.00936}}].

\bibitem{Kallosh:2015lwa}
R.~Kallosh and A.~Linde, \emph{{Planck, LHC, and $\alpha$-attractors}}, \href{https://doi.org/10.1103/PhysRevD.91.083528}{\emph{Phys. Rev. D} {\bfseries 91} (2015) 083528} [\href{https://arxiv.org/abs/1502.07733}{{\ttfamily 1502.07733}}].

\bibitem{Artymowski:2016pjz}
M.~Artymowski and J.~Rubio, \emph{{Endlessly flat scalar potentials and $\alpha$-attractors}}, \href{https://doi.org/10.1016/j.physletb.2016.08.024}{\emph{Phys. Lett. B} {\bfseries 761} (2016) 111} [\href{https://arxiv.org/abs/1607.00398}{{\ttfamily 1607.00398}}].

\bibitem{Figueroa:2015rqa}
D.G.~Figueroa, J.~Garcia-Bellido and F.~Torrenti, \emph{{Decay of the standard model Higgs field after inflation}}, \href{https://doi.org/10.1103/PhysRevD.92.083511}{\emph{Phys. Rev. D} {\bfseries 92} (2015) 083511} [\href{https://arxiv.org/abs/1504.04600}{{\ttfamily 1504.04600}}].

\bibitem{Enqvist:2015sua}
K.~Enqvist, S.~Nurmi, S.~Rusak and D.~Weir, \emph{{Lattice Calculation of the Decay of Primordial Higgs Condensate}}, \href{https://doi.org/10.1088/1475-7516/2016/02/057}{\emph{JCAP} {\bfseries 02} (2016) 057} [\href{https://arxiv.org/abs/1506.06895}{{\ttfamily 1506.06895}}].

\bibitem{DiscreteDerivatives}
B.~Fornberg, \emph{{Generation of Finite Difference Formulas on Arbitrarily Spaced Grids}}, \href{https://doi.org/10.1103/PhysRevE.62.1368}{\emph{Mathematics of Computation 51} {\bfseries 184} (1988) 699}.

\end{thebibliography}\endgroup
%%%%%%%%%%%%%%%%%%%%%%%%%%%%%%%%%%%%%%%%%%%%%%%%%%%%%%%%%%%%%%%%%%%%%%%%%%%%%%%%
\end{document}